\documentclass[final,usenatbib,changeblnc]{iupartex-v23}  
\usepackage{newtxtext,newtxmath}
\usepackage[T1]{fontenc}
\usepackage{ae,aecompl}

\usepackage{multicol}   
\usepackage{bm}		    

\title[Multiple System EM Boo]{Spectroscopic Disentangling Revealed the Tertiary Component in the Multiple System EM\,Boo}

\author[Yıldız et al.]{Ö. H. Yıldız$^{1}$\orcid{0000-0002-5078-1293}, H. Bakış$^{2}$\orcid{0000-0001-8875-5464}, and B. Özkardeş$^{3}$\orcid{0000-0002-6764-9299}
\affsep \\
$^1$Akdeniz University, Graduate Institute of Natural and Applied Sciences, Antalya, Türkiye\\
$^2$Akdeniz University, Faculty of Science, Space Sciences and Technologies Department, Antalya, Türkiye\\
$^3$Çanakkale Onsekiz Mart University, Faculty of Science, Space Sciences and Technologies Department, Çanakkale, Türkiye
}

\corres{Hicran Bakış}{hicranbakis@akdeniz.edu.tr}

\pubyear{2026}

\doiheader{XXXXXXX/PAR.20XX.00000}
\date{
	\pSubmit{02.04.2026} 
	\pRevReq{22.04.2026}
	\pLastRevRec{27.04.2026}
	\pAccept{04.05.2026}
	\pPubOnl{XX.05.2026}
}
\volume{4}
\volnumber{1}

\begin{document}
\label{firstpage}
\pagerange{\pageref*{firstpage}--\pageref*{lastpage}}
\maketitle

\begin{abstract}
We present a comprehensive photometric and spectroscopic study of the triple stellar system EM\,Boo. The system is composed of detached, low-mass components, and for the first time in the literature, the spectrum of the tertiary component has been successfully disentangled from the composite spectrum using the \texttt{KOREL} code. Synthetic spectra were generated for each disentangled component, allowing determination of their atmospheric parameters. The depth of the H$_\alpha$ line in the tertiary spectrum indicates that it is an intermediate-temperature star, consistent with spectral types between A and F, and its effective temperature was determined to be 7000~K. By analyzing the radial velocity and light curves simultaneously, the fundamental physical parameters of the system were derived, and its detailed evolutionary status was investigated using \texttt{MESA} models. The \textit{HIPPARCOS} trigonometric parallax ($\varpi_{\rm Hip}=1.33\pm1.45$ mas) and \textit{Gaia} DR3 trigonometric parallax ($\varpi_{\rm Gaia}=3.9699\pm0.1812$ mas) show a significant discrepancy, most likely related to the system’s multiplicity and the limitations of single-star astrometric solutions. To provide independent distance estimates, we modeled the spectral energy distribution (SED) using multi-wavelength flux data, yielding $E(B-V)=0.05$ mag and a trigonometric parallax $\varpi_{\rm SED}=3.2$ mas, corresponding to $d_{\rm SED}=313$ pc. Furthermore, photometric distance estimates based on the components’ absolute magnitudes yield $d_{1}=299$ pc and $d_{2}=301$ pc, in good agreement with the SED-based distance. Both the SED-based and photometric distances converge around $d=300$ pc, indicating that the \textit{Gaia} trigonometric parallax underestimates the true distance of EM\,Boo.
\end{abstract}


\section{Introduction}\label{sec:intro}
Detached eclipsing binary systems play a crucial role in stellar astrophysics, as they provide one of the most reliable means of determining fundamental stellar parameters such as masses, radii, and effective temperatures with high precision. When combined with spectroscopic observations, these systems serve as benchmark objects for testing stellar structure and evolutionary models. In particular, detached Algol-type binaries are valuable laboratories for studying stellar evolution in close systems before significant mass transfer.

EM\,Boo (HD 130617 = HIP 72426, $\alpha = 14^\text{h} 48^\text{m} 32^\text{s}.2333$, $\delta = +24^\circ 45' 03''.7993$) is a detached Algol-type eclipsing binary system located in the northern hemisphere. EM\,Boo constitutes the primary binary component of the multiple stellar system, the Washington Double Star Catalog (WDS) J14485+2445AB. It was first identified as a visual binary by \citet{couteau1970}, who observed the system with a 50 cm telescope and cataloged it as Cou 304. Subsequent observations by the HIgh Precision PARallax COllecting Satellite \citep[\textit{HIPPARCOS;}][]{Hip1997} provided precise astrometric and photometric measurements, including trigonometric parallax, colour indices, and photometry in the $H_{\rm p}=9.135\pm 0.003
$, $B_{\rm T}=9.659 \pm 0.016$, and $V_{\rm T}=9.119 \pm 0.016$ mag. According to the \textit{HIPPARCOS} data, the brightness difference between the eclipsing pair and the third component is $\Delta m=2.17\pm 0.07$ mag. The angular separation between the components is $\rho=0.47$ arcseconds, with a position angle of $\theta=306^\circ$. 

In addition, WDS reports astrometric parameters at two epochs: $\rho = 0.5$ arcseconds, $\theta = 315^\circ$ (1968) and $\rho = 0.4$ arcseconds, $\theta = 297^\circ$ (2014). The catalog also lists component magnitudes of $9.27$ mag for A and $11.44$ mag for B, corresponding to a magnitude difference of $2.17$ mag between the eclipsing binary and the tertiary component. This photometric contrast provides the basis for estimating the third-light contribution in subsequent analyses.

 We note that the identification of component A as a close pair arises from subsequent eclipsing-binary classification and is not implied by the WDS photometric entries. Notably, the \textit{HIPPARCOS} observations in 1992 fall almost midway between the two WDS epochs (1968 and 2014). This temporal placement suggests that the \textit{HIPPARCOS} measurements may reflect the same gradual change in separation and position angle reported in the WDS catalog, supporting the evidence for a consistent astrometric trend over several decades.

The physical association of the third component with the eclipsing binary remains uncertain. This ambiguity makes EM\,Boo a particularly interesting case for investigating the role of 
tertiary companions in binary evolution, as the contribution of third light can significantly affect photometric and spectroscopic analyses. This makes EM\,Boo not only a benchmark detached Algol-type binary, but also a rare case for testing the influence of tertiary companions on binary evolution.

EM\,Boo was later included in binary star catalogs based on \textit{HIPPARCOS} and \textit{TYCHO-2} data  \citep{Fabricius2000, Fabricius2002}. The system was first recognized as an eclipsing binary by \citet{Malkov2006}. Subsequently, \citet{Szczygie2008} listed EM\,Boo in a catalog of binaries exhibiting coronal activity, reporting its $V$- and $I$-band magnitudes, as well as its orbital period and distance.

The effective temperature of the EM\,Boo system was initially estimated as 5901 K by \citet{McDonald2012}, who also derived a distance of 752 pc. This value was later revised to 6023 K by \citet{Bermejo2013} using a principal component analysis (PCA) approach. The most recent detailed study of the system is the radial-velocity analysis by \citet{Ozkardes2018}, who derived semi-amplitudes of $K_1 = 100.7 \pm 2.6$ km s$^{-1}$ and $K_2 = 120.1 \pm 2.6$ km s$^{-1}$ for the primary and secondary components, respectively. Despite these efforts, no comprehensive study combining modern photometric, spectroscopic, and atmospheric analyses of the EM\,Boo system has been presented to date.

In this work, we present the first comprehensive investigation of EM\,Boo that simultaneously models radial-velocity measurements, multi-band photometric light curves, spectral disentangling, atmospheric parameters, and evolutionary tracks within a single framework. In addition, the explicit contribution of the third light is taken into account, while the possible physical association of the tertiary component is discussed. This makes EM\,Boo an important addition to the sample of well-characterized detached eclipsing binaries and a rare case for exploring the influence of tertiary companions.

The paper is organized as follows: Section~\ref{sec:section2} describes the observations and data reduction procedures. Radial velocity measurements and the orbital solution are presented in Section~\ref{sec:section3}. The light-curve analysis is given in Section~\ref{sec:section4}, followed by atmospheric modeling in Section~\ref{sec:section5}. The astrophysical parameters and the distance of the system are detailed in Section~\ref{sec:section6}, while evolutionary modeling is covered in Section~\ref{sec:evolutionary_status}. Finally, Section~\ref{sec:conclusions}  discusses the results in the context of previous studies and summarizes the main conclusions.

\section{Observations and Data Reduction}
\label{sec:section2}

\begin{figure}[!t]
    \centering
    \includegraphics[width=\linewidth]{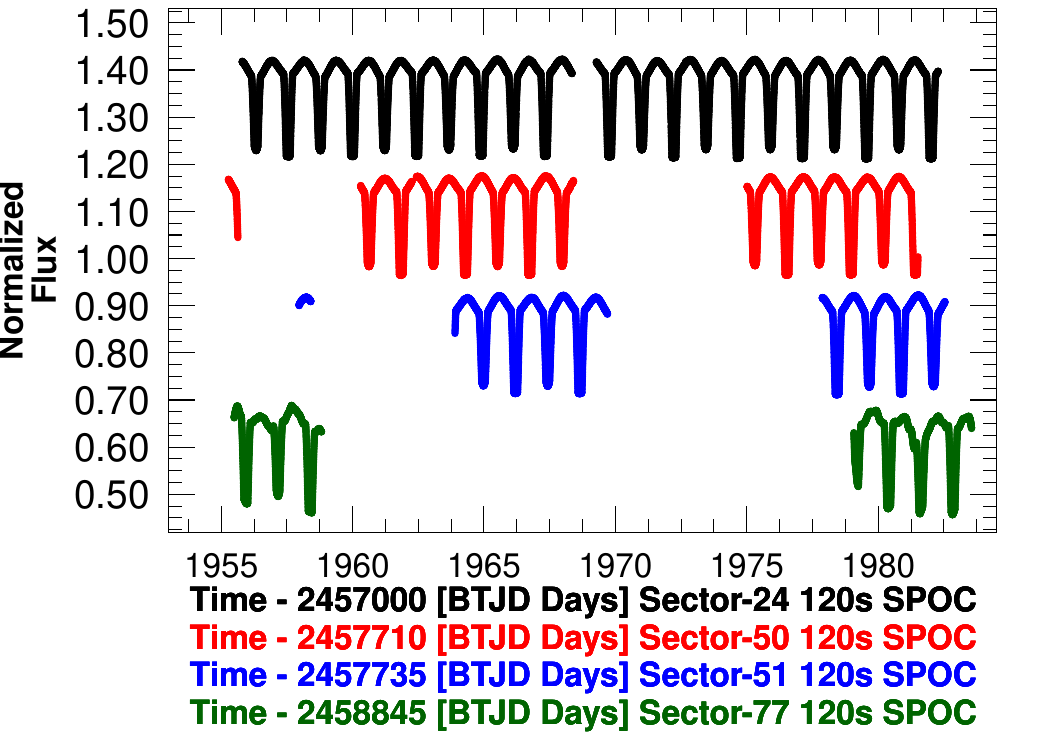}
    \caption{Available photometric \textit{TESS} data for EM\,Boo with the quality mask ``hard''.}
    \label{fig:tess_data}
\end{figure}

Photometric observations of the eclipsing binary system EM\,Boo were carried out using the 25-cm telescope (AUT25) at the Department of Space Sciences and Technologies, Akdeniz University. The telescope is equipped with a QSI-632ws CCD camera and standard Johnson $BVRI$ filters. In addition to the ground-based observations, EM\,Boo was also observed by the Transiting Exoplanet Survey Satellite \citep[\textit{TESS,}][]{Ricker2015}. \textit{TESS} has observed EM\,Boo in four different sectors, 24, 50, 51, and 77. Among these sectors, we have chosen the data reduced with \texttt{SPOC} products with 120-second exposure times (see Figure~\ref{fig:tess_data}). The photometric data were retrieved from the Mikulski Archive for Space Telescopes (MAST)\footnote{\url{https://mast.stsci.edu/}} via using \texttt{lightkurve v2.5} \citep{lightkurve} with the quality setting of ``hard''. In our analysis, we have used data from sector 24, since it is the only sector that is continuous and fully covered. \texttt{SAP} flux was specifically selected to see if there are some long term effects\footnote{\url{https://heasarc.gsfc.nasa.gov/docs/tess/LightCurveFile-Object-Tutorial.html}}. Normalization of the \textit{TESS} data has been done internally via \texttt{lightkurve}.

Spectroscopic observations were performed with the 60-cm telescope (UBT60) at the Department of Space Sciences and Technologies, Akdeniz University, equipped with the \textsc{SHELLYAK} echelle spectrograph \citep{Ozkardes2018}. The UBT60 is an alt-azimuth-mounted telescope, while the \textsc{SHELLYAK} spectrograph provides a resolving power of $R \approx 12,000$ and records 27 echelle orders covering the wavelength range from 405 to 816 nm. Using this setup, a total of 16 spectra of the EM\,Boo system were obtained. The details of these spectroscopic observations are summarized in Table~\ref{tab:sec1}.

The spectroscopic data reduction was carried out using the Image Reduction and Analysis Facility (\texttt{IRAF}) software package, following standard procedures. These include bias subtraction, flat-field correction, and cosmic-ray removal. Wavelength calibration was performed using comparison lamp spectra, after which the spectra were normalized to the continuum level. These steps ensured that the final reduced spectra were suitable for subsequent radial-velocity measurements and spectral analysis.

In addition to our own spectroscopic observations, EM\,Boo was also observed with the ELODIE spectrograph \citep{Baranne1996}, a high-resolution, fiber-fed echelle instrument with a resolving power of $R \sim 42,000$, formerly mounted on the 1.93-m telescope at the Observatoire de Haute-Provence (OHP). A total of 17 archival ELODIE spectra were retrieved from the public OHP archive \citep{Moultaka2004}\footnote{\url{https://atlas.obs-hp.fr/elodie/}}. The details of the ELODIE observations are also listed in Table~\ref{tab:sec1}.

The spectroscopic data serve multiple purposes in this study. In addition to providing the radial velocities required for the orbital solution, the disentangled spectra are used for atmospheric modeling of the individual components. The photometric light curves, combined with the radial-velocity measurements, allow us to determine the absolute physical parameters of the system and its components in a consistent framework that simultaneously accounts for photometric and spectroscopic constraints. These fundamental parameters are subsequently used as inputs for evolutionary modeling of both stars, carried out using the Modules for Experiments in Stellar Astrophysics
(\texttt{MESA}) stellar evolution code \citep{Paxton2011}.
\begin{table}[!t]
\setlength{\tabcolsep}{1.4pt}
\renewcommand{\arraystretch}{1.0}
\small
    \centering
    \caption{Log of radial-velocity measurements of EM\,Boo, including orbital phases, residuals, exposure times, and source.}
    \label{tab:sec1}
    \begin{tabular}{cccrrrrc}
        \hline
        HJD (-2400000) & Exp.Time & Phase & RV$_1$ & (O-C)$_1$ &  RV$_2$ & (O-C)$_2$& Ref \\
        (days) & (s) &   & \multicolumn{2}{c}{(km~s$^{-1}$)} & \multicolumn{2}{c}{(km~s$^{-1}$)} &  \\[-0.9ex]

        \hline
        51406.3312 & 1503.6 & 0.628 & 62.8  & 5.6  & -101.5 & -1.3& 1 \\
        51409.3454 & 1889.2 & 0.861 & 61.3  & -0.4 & -103.9 & 1.8& 1  \\
        51740.3653 & 1800.7 & 0.178 & -99.5 & 2.6  & 109.1  & 14.9& 1 \\
        51741.3470 & 2401.9 & 0.580 & 34.2  & 0.7  & -71.8  & -0.6& 1 \\
        51742.3436 & 2400.9 & 0.987 & 9.9   & 15.7 & -19.6  & 3.8& 1  \\
        51927.6816 & 1800.7 & 0.752 & 75.6  & -8.9 & -135.1 & -1.6& 1 \\
        51929.6667 & 1490.6 & 0.563 & 16.9  & -7.3 & -61.6  & -1.7& 1 \\
        51930.6656 & 1800.8 & 0.971 & 0.4   & -3.4 & -35.2  & -0.1& 1 \\
        51931.6343 & 1795.6 & 0.367 & -94.8 & -8.4 & 72.3   & -2.6& 1 \\
        52041.5802 & 1800.7 & 0.312 & -106.5& -2.0 & 96.2   & -0.8& 1 \\
        52042.5089 & 1800.7 & 0.692 & 82.3  & 4.3  & -129.2 & -3.6 & 1\\
        52299.6046 & 2000.8 & 0.790 & 77.6  & -3.8 & -132.4 & -2.7& 1 \\
        52300.6411 & 1600.7 & 0.214 & -114.8& -5.4 & 105.0  & 2.0& 1  \\
        52489.3564 & 1700.8 & 0.359 & -86.8 & 3.0  & 74.1   & -5.0& 1 \\
        52491.3584 & 2400.9 & 0.151 & -93.9 & -0.3 & 85.1   & 1.3& 1  \\
        57871.2736 & 600  & 0.436 & -49.1 & 2.8  & 35.3   & 2.4&2  \\
        57871.3734 & 600  & 0.477 & -24.8 & 2.9  & 3.9    & 0.5 & 2 \\
        57871.3990 & 600  & 0.488 & -24.1 & -2.8 & -10.4  & -6.0& 2 \\
        57871.4114 & 600  & 0.493 & -23.2 & -5.0 & -3.8   & 4.4& 2  \\
        57872.2606 & 600  & 0.840 & 70.6  & 1.3  & -117.4 & -2.5& 2 \\
        57910.5407 & 600  & 0.488 & -24.5 & -3.6 & -5.8   & -0.8& 2 \\
        57911.4918 & 600  & 0.877 & 59.1  & 4.3  & -92.7  & 4.5& 2  \\
        57912.4433 & 600  & 0.266 & -113.8& -2.4 & 105.6  & 0.1& 2  \\
        57916.4011 & 600  & 0.884 & 52.8  & 1.1  & -89.5  & 3.9& 2  \\
        57917.3906 & 600  & 0.289 & -111.6& -2.6 & 102.0  & -0.6& 2 \\
        57927.4136 & 600  & 0.386 & -74.7 & 3.5  & 58.7   & -6.3 & 2\\
        57934.4031 & 600  & 0.243 & -112.3& -0.5 & 105.7  & -0.3 & 2\\
        57936.4026 & 600  & 0.060 & -48.8 & 1.3  & 33.2   & 2.5 & 2 \\
        57939.3997 & 600  & 0.286 & -111.8& -2.4 & 98.7   & -4.4& 2 \\
        57944.4077 & 600  & 0.333 & -97.2 & 1.7  & 94.1   & 3.9& 2  \\
        57946.4097 & 600  & 0.177 & -93.9 & 8.0  & 85.6   & -8.2 & 2\\
        \hline
        \multicolumn{8}{l}{Note: (1) ELODIE, and (2) UBT60}
    \end{tabular}
\end{table}
\section{Radial Velocities Measurements And Orbital Solutions}
\label{sec:section3}

 The EM\,Boo system has been observed at different epochs with two spectrographs. The spectra obtained at the UBT Observatory were previously analysed by \citet{Ozkardes2018} using the \texttt{KOREL} code. In the present study, we extend this analysis by incorporating additional archival spectra obtained with the ELODIE spectrograph. Using the combined dataset, we repeated the spectral disentangling and radial-velocity determination with \texttt{KOREL} in order to improve both the quality of the disentangled component spectra and the precision of the derived orbital parameters.

Since the accuracy of spectral disentangling and radial-velocity measurements increases with the number of available spectra, the combined dataset was analysed independently in two spectral regions, namely H$_\alpha$ and H$_\beta$. The \texttt{KOREL} code was employed to measure the radial velocities and determine the spectroscopic orbital parameters, being particularly well suited for analysing shallow and partially blended spectral lines. Initial values of the orbital parameters were adopted from \citet{Ozkardes2018}. The time of minimum ($T_0$), eccentricity ($e$), longitude of periastron, and the radial-velocity semi-amplitudes of the primary and secondary components ($K_1$ and $K_2$) were treated as free parameters in the analysis. Although the eccentricity was initially allowed to vary, the solution was consistent with a circular orbit, and $e$ was subsequently fixed to zero.

The orbital period of the system was refined by combining previously published times of minima with newly determined minima derived from the \textit{TESS} photometry. A least-squares fit to all available minima yielded the following revised ephemeris:\vspace*{6pt}
\begin{equation}
\small
\mathrm{Min~I~(BJD_{TDB})} = 2460424.0705(2) + E \times 2.44623996(84),\vspace*{4pt}
\end{equation}
which was subsequently fixed in the spectroscopic analysis.

Both the photometric and spectroscopic datasets indicate the presence of a third-light contribution in the system. Accordingly, this effect was taken into account consistently in the analysis of both datasets. As the \texttt{KOREL} code does not determine the systemic (centre-of-mass) velocity of the system, the radial velocities obtained from the disentangling procedure are relative to an arbitrary zero point. To place the measurements on an absolute scale, we adopted a systemic velocity of $V_\gamma = -14.6 \pm 3.1$ km s$^{-1}$ from \citet{Ozkardes2018}.

\begin{figure}[!t]
    \centering
    \includegraphics[width=0.99\linewidth]{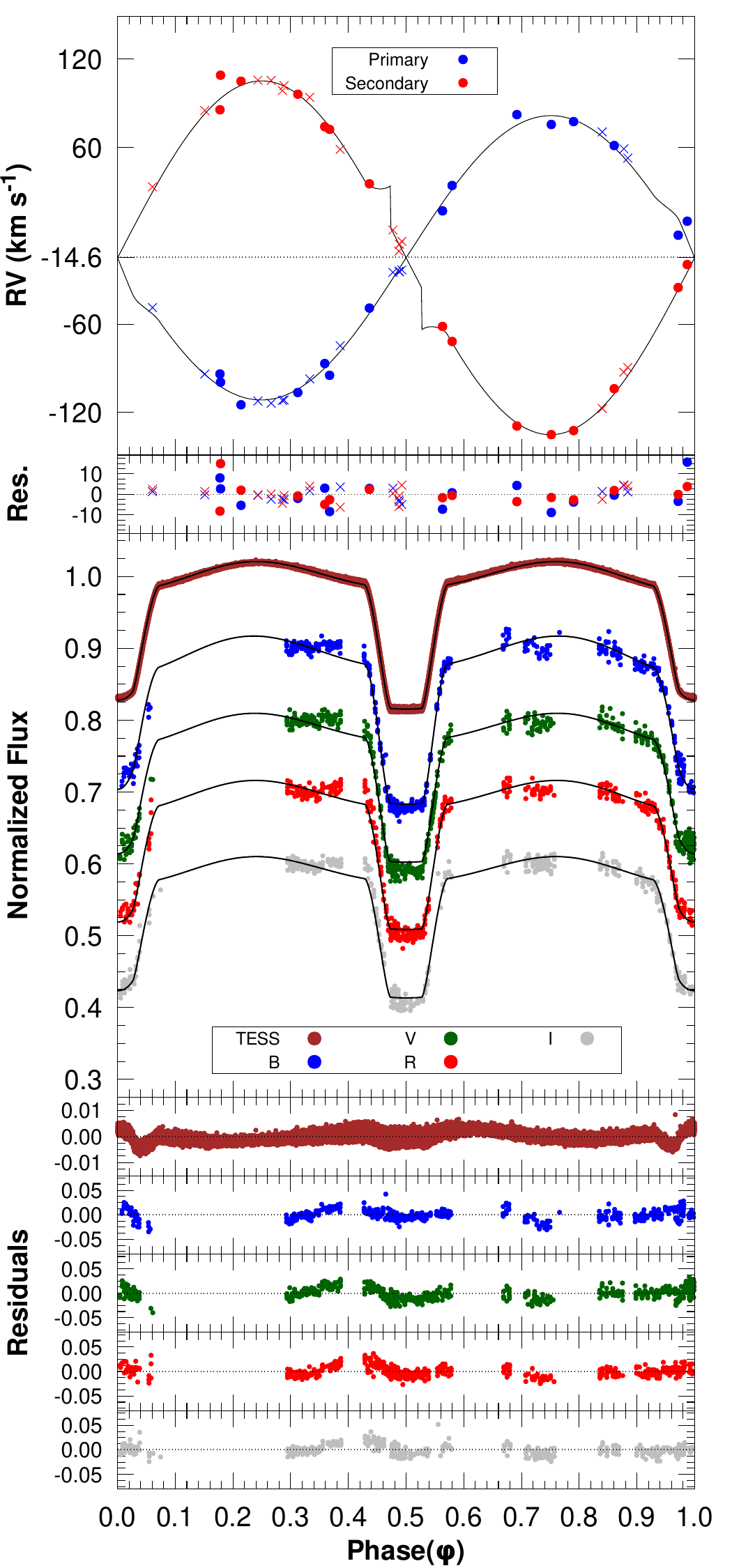}
    \caption{Combined radial-velocity and light-curve solutions for EM\,Boo. The upper panels show the observed radial velocities of both components with the fitted orbital model and residuals. The radial velocities that have been obtained from each source showed in different symbols, filled circles and cross, for ELODIE and UBT60, respectively. The lower panels present the $\textit{TESS}$ and Johnson $BVRI$ light curves with their best-fitting models and residuals as a function of orbital phase.}
    \label{fig:embooph}
\end{figure}

For each observing epoch, the radial velocities derived from the H$_\alpha$ and H$_\beta$ regions were combined using weighted means based on their respective O–C residuals. The resulting radial velocities, together with the exposure times and their residuals from the adopted model, are listed in Table~\ref{tab:sec1}. The radial-velocity curves of the primary and secondary components are shown in Figure~\ref{fig:embooph}. Assuming a circular orbit, the final spectroscopic solution yields semi-amplitudes of $K_1 = 98.3 \pm 1.3$ km s$^{-1}$ and $K_2 = 120.3 \pm 1.1$ km s$^{-1}$ for the primary and secondary components, respectively.

The spectroscopic orbital parameters derived from this analysis, along with the component masses and orbital separation inferred from them, are summarized in Table~\ref{tab:solution}.
    
    

\section{Simultaneous Analysis of Light and Radial Velocity Curves}
\label{sec:section4}
The EM\,Boo system is classified as a detached eclipsing binary of late spectral type. Although one of the components is nearly filling its Roche lobe, mass transfer has not yet begun in the system. 

The photometric data obtained with the AUT25 telescope, together with those from the $\textit{TESS}$ mission, were analyzed simultaneously with the radial velocity data for light curve modeling using the Wilson–Devinney ($\texttt{WD}$) code \citep{Wilson1994, Wilson2008, Wilson2012, 2020Wilson} through the graphical interface $\texttt{PHOEBE}$ v1.0 \citep{Prsa2005}. In the analysis, the detached binary configuration was adopted. \citet{Bermejo2013} and \citet{McDonald2012} reported the effective temperature of the system as 6023 K and 5901 K, respectively; however, these values were not provided separately for the individual components. Since the system does not contain early-type components, the Q-method based on photometric colours cannot be applied to derive the temperatures. Instead, the component temperatures can be determined either by assuming that the reported values correspond to the secondary star, or more reliably, by applying model atmospheres to disentangled spectra. Our analysis was carried out in an iterative manner. In the first step, an initial light-curve solution was obtained by fixing the effective temperature of one component to the value adopted from the literature. Since the light-curve solution requires the secondary star to be the hotter of the two components, the literature temperature was initially assigned to the secondary component. This preliminary solution was then used to determine the light contributions of both components, which were applied in the spectral disentangling with $\texttt{KOREL}$ code to obtain the individual component spectra.

Atmospheric model fitting was subsequently performed on the disentangled spectra to derive updated effective temperatures for both components. In the final step, these spectroscopically determined temperatures were adopted in a new light-curve analysis, yielding the final and self-consistent set of light-curve parameters.

In this model, the fixed parameters were the orbital period ($P$) and, in the final iteration, the effective temperature of the primary component ($T_1$), while the free parameters were the time of minimum ($T_0$), the surface potentials ($\Omega_{1,2}$), the luminosity of the primary ($L_1$), and the orbital inclination ($i$). Bolometric albedos ($A_{1,2}=0.5$) and gravity-darkening coefficients ($g_{1,2}=0.32$), appropriate for convective envelopes, were adopted \citep{Lucy1967, Rucinski1969}. Limb-darkening coefficients for each filter were taken from \citet{1993Vanhamme}. The mass ratio ($q$) was treated as a free parameter and determined from a simultaneous solution of the radial-velocity and light curves.

An additional important parameter is the third-light contribution. This effect was treated as a free parameter in the solution, ensuring that the light-curve modeling consistently accounts for the presence of the tertiary component. Based on the \textit{HIPPARCOS} magnitude difference discussed in Section~\ref{sec:intro}, the third-light contribution was estimated to be $\sim 14-15\%$. Given this relatively low light contribution, the direct detection of the tertiary component in the composite spectra is expected to be challenging.

\begin{figure*}[!t]
    \centering
    \includegraphics[width=0.49\linewidth]{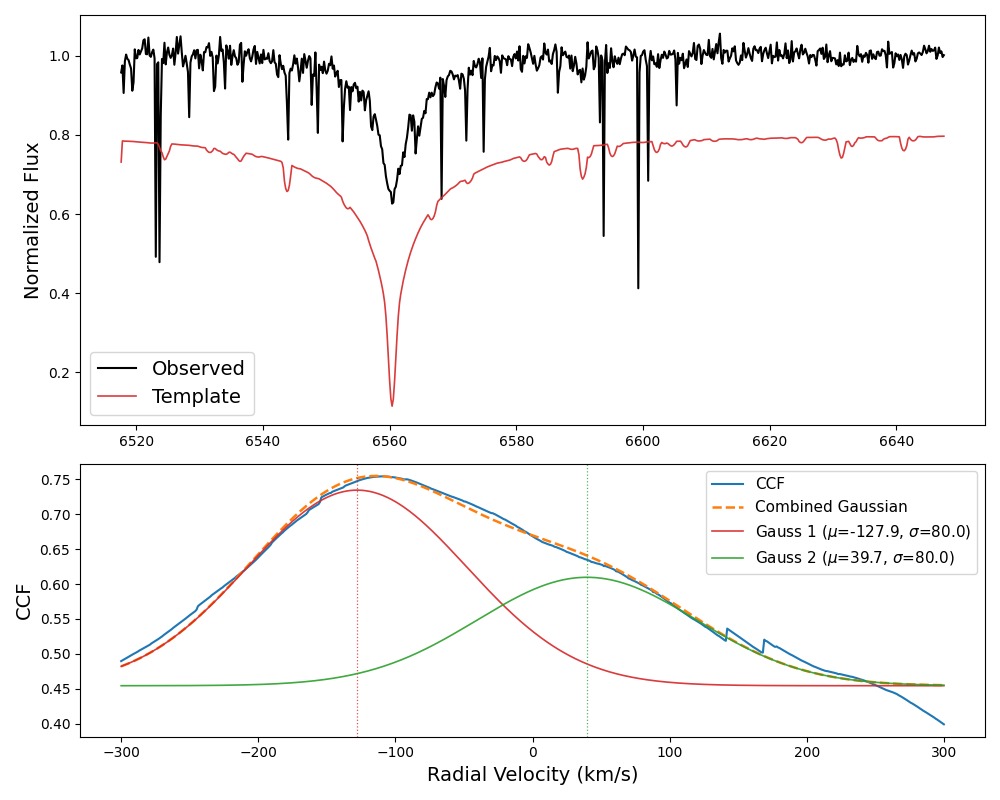}\hspace{2pt}
    \includegraphics[width=0.49\linewidth]{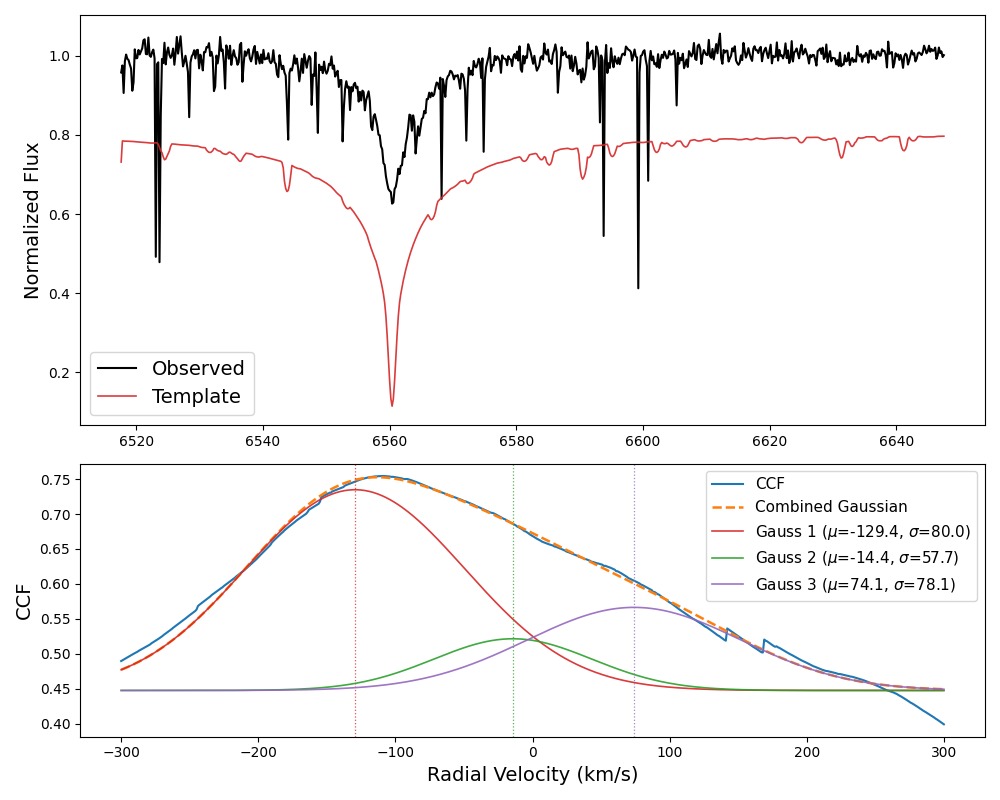}
    \caption{Cross-correlation of observed spectrum with synthetic spectrum of the third component (top panels) and resulting CCF functions with two-(left) and three-gaussian (right) fittings.}
    \label{fig:ccf}
\end{figure*}

To demonstrate the detectability of the tertiary component, we performed a cross-correlation analysis at orbital phase $\sim$0.25, where the spectral lines of the close binary components are maximally separated (see Figure~\ref{fig:ccf}). The resulting CCF profile was modeled using both two- and three-Gaussian components. Although the three-Gaussian model yields a better fit, the signature of the tertiary component remains weak and cannot be clearly distinguished by eye. This highlights the limitation of single-spectrum diagnostics and justifies the use of spectral disentangling, which exploits the full set of observed spectra.
\begin{table*}[!t]
	\setlength{\tabcolsep}{10pt}
	\renewcommand{\arraystretch}{1.4}
\centering
\caption{Parameters determined for EM\,Boo from the analysis of RV and LC data.} 
\label{tab:absolutepar}
\begin{tabular}{lccc}\hline\hline\noalign{\vspace*{-4pt}}
Parameter                       & Symbol                                & \multicolumn{2}{c}{This Study} \\[-1.5ex]
& & Primary                       & Secondary  \\[-1ex]
\hline
Separation ($R_\odot$)          & $a$                                   & \multicolumn{2}{c}{$10.6^{+0.2}_{-0.2}$}               \\
Mass ratio                      & \emph{q}                              & \multicolumn{2}{c}{$0.81^{+0.2}_{-0.2}$}               \\
Systemic velocity (km s$^{-1}$) & \(V_\gamma\)                          & \multicolumn{2}{c}{$-14.6\pm 3.1$}              \\
Eccentricity                    & \emph{e}                              & \multicolumn{2}{c}{0 (fixed)}                               \\
Orbital inclination ($^\circ$)  & \emph{i}                              & \multicolumn{2}{c}{$89.186^{+0.059}_{-0.025}$}              \\
Temperature ratio               & $T_\mathrm{eff,2}/T_\mathrm{eff,1}$   & \multicolumn{2}{c}{$1.057^{+0.001}_{-0.001}$}               \\
Mass ($M_\odot$)                & \emph{M}                              & $1.476^{+0.103}_{-0.097}$ & $1.195^{+0.086}_{-0.082}$ \\
Radius ($R_\odot$)              & \emph{R}                              & $3.213^{+0.065}_{-0.064}$    & $1.407^{+0.032}_{-0.030}$    \\
Surface gravity (cgs)           & $\log g$                              & $3.593^{+0.047}_{-0.047}$    & $4.219^{+0.049}_{-0.050}$    \\
Light ratio (\textit{TESS})     & $l/l_\mathrm{total}$                  & $0.701^{+0.003}_{-0.003}$    & $0.154^{+0.003}_{-0.003}$ \\
\hline\noalign{\vspace*{-3pt}}
Light ratio of the third body (\textit{TESS})     & $l_{3}/l_\mathrm{total}$& \multicolumn{2}{c}{$0.145^{+0.004}_{-0.003}$} \\
\hline
\hline
\label{tab:solution}
\end{tabular}\vspace*{-20pt}
\end{table*}

The light contributions obtained from this solution were subsequently used to disentangle the composite spectra of EM\,Boo into the individual component spectra. Owing to its significantly larger light contribution, the primary component yielded a higher-quality disentangled spectrum, allowing for a more reliable atmospheric model fitting. Consequently, the effective temperature of the primary component was determined to be $T_1 = 6000$ K and was fixed in the final light-curve solution to ensure a stable and physically consistent refinement. The effective temperature derived for the secondary component from the final light-curve solution was found to be consistent with the value obtained independently from the atmospheric modeling of its disentangled spectrum, confirming the reliability of the adopted methodology and the final system parameters.

Parameter uncertainties were estimated using a Markov Chain Monte Carlo (MCMC) analysis implemented with the \texttt{emcee} sampler \citep{Foreman-Mackey2013}, employing 128 walkers and 1000 steps. Due to the insufficient phase coverage of the ground-based \textit{BVRI} light curves, the MCMC analysis was applied exclusively to the \textit{TESS} light curve, which provides continuous and homogeneous phase coverage over the entire orbital cycle. The \textit{TESS} light curves do not also show a significant or persistent asymmetry between maxima, except for a marginal deviation in Sector 77 with limited data coverage. The residuals in LC do not exhibit a clear systematic trend with orbital phase. Although stellar activity (e.g., starspots) is physically plausible for stars with convective envelopes, test spot models were not found to provide a meaningful or statistically significant improvement to the fit. Therefore, we adopted a spot-free model.

\begin{figure}[!h]
    \centering
    \includegraphics[width=0.99\linewidth]{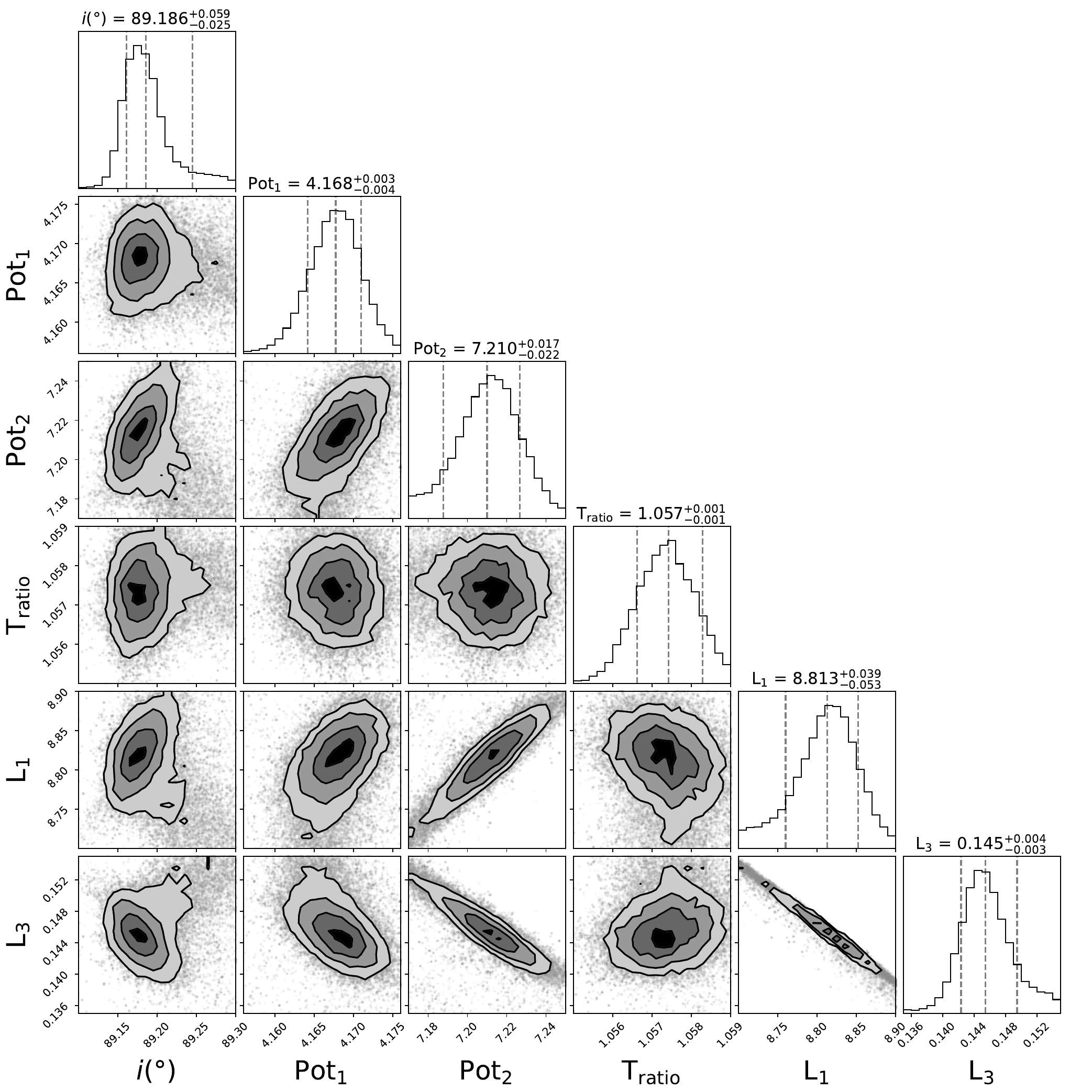}
    \caption{Posterior distributions of selected system parameters derived from the MCMC analysis of the \textit{TESS} light curve. The diagonal panels show the one-dimensional marginalized distributions, and the off-diagonal panels illustrate parameter correlations through joint posterior probability densities. The contours represent the 1$\sigma$ and 2$\sigma$ confidence levels.}
    \label{fig:mcmc}
\end{figure}

The photometric and radial-velocity data, together with the best-fitting model curves obtained from the final solution, are shown in Figure~\ref{fig:embooph}, including the corresponding residuals. To investigate possible zero-point offsets between datasets obtained with two different sources given in Table~\ref{tab:sec1}, we examined the O–C residuals of the radial velocities by distinguishing the UBT60 and ELODIE measurements for both components. No systematic deviation between the datasets is observed, indicating that any instrumental zero-point differences are negligible within the measurement uncertainties. Therefore, the adopted systemic velocity does not introduce a significant bias in the orbital solution. The posterior probability distributions of selected parameters derived from the MCMC analysis, with \textit{TESS} data, along with their multivariate Gaussian representation, are presented in Figure~\ref{fig:mcmc}. The results of the simultaneous radial-velocity and light-curve analyses, together with the associated parameter uncertainties derived from the MCMC analysis, are summarized in Table~\ref{tab:solution}.

Since the MCMC method was not applied to the \textit{BVRI} light curves, uncertainties for parameters derived from these data sets were not estimated in the same manner. Instead, the light contributions obtained from the \textit{BVRI} solutions are presented separately in Table ~\ref{tab:lightcont}. We note that the \textit{BVRI} light-curve fits are fully consistent with the parameters derived from the \textit{TESS}-based solution, confirming the reliability and internal consistency of the adopted model across all photometric data sets.

\begin{table}[!t]
	\setlength{\tabcolsep}{10pt}
	\renewcommand{\arraystretch}{1.1}
\centering
\caption{Light contributions of the primary, secondary, and tertiary components in different photometric passbands. The contributions are normalized to the total system flux, such that $L_1 + L_2 + L_3 = 1$.}
\label{tab:light_contributions}
\begin{tabular}{lccc}
\hline
Filter & Primary & Secondary & Tertiary \\
\hline
\textit{B}    & $0.675 $ & $0.175   $ & $0.150   $ \\
\textit{V}    & $0.664   $ & $0.162   $ & $0.174   $ \\
\textit{R}    & $0.697   $ & $0.158   $ & $0.145   $ \\
\textit{I}    & $0.687   $ & $0.155   $ & $0.158   $ \\
\textit{TESS} & $0.701   $ & $0.154   $ & $0.145   $ \\
\hline
\label{tab:lightcont}
\end{tabular}\vspace*{-10pt}
\end{table}
\section{Atmospheric Modeling of the Disentangled Spectra}\label{sec:section5}

EM\,Boo light curve analysis was carried out using ground-based Johnson \textit{BVRI} photometry together with the $\textit{TESS}$ light curve. As a result, the relative light contributions of the individual components are known as a function of orbital phase over a broad wavelength range. These light ratios were used as input for the spectral disentangling procedure, allowing the extraction of the individual component spectra using the $\texttt{KOREL}$ code. This approach has been successfully applied to several binary systems (e.g., $\delta$ Lib; \citet{2006Bakis}, V716 Cep; \citet{2008Bakis}, MQ Cen; \citet{2019Bakis}, RS Sgr; \citet{2025Bakis}).

\begin{figure*}[!t]
    \centering
    \includegraphics[width=0.85\linewidth]{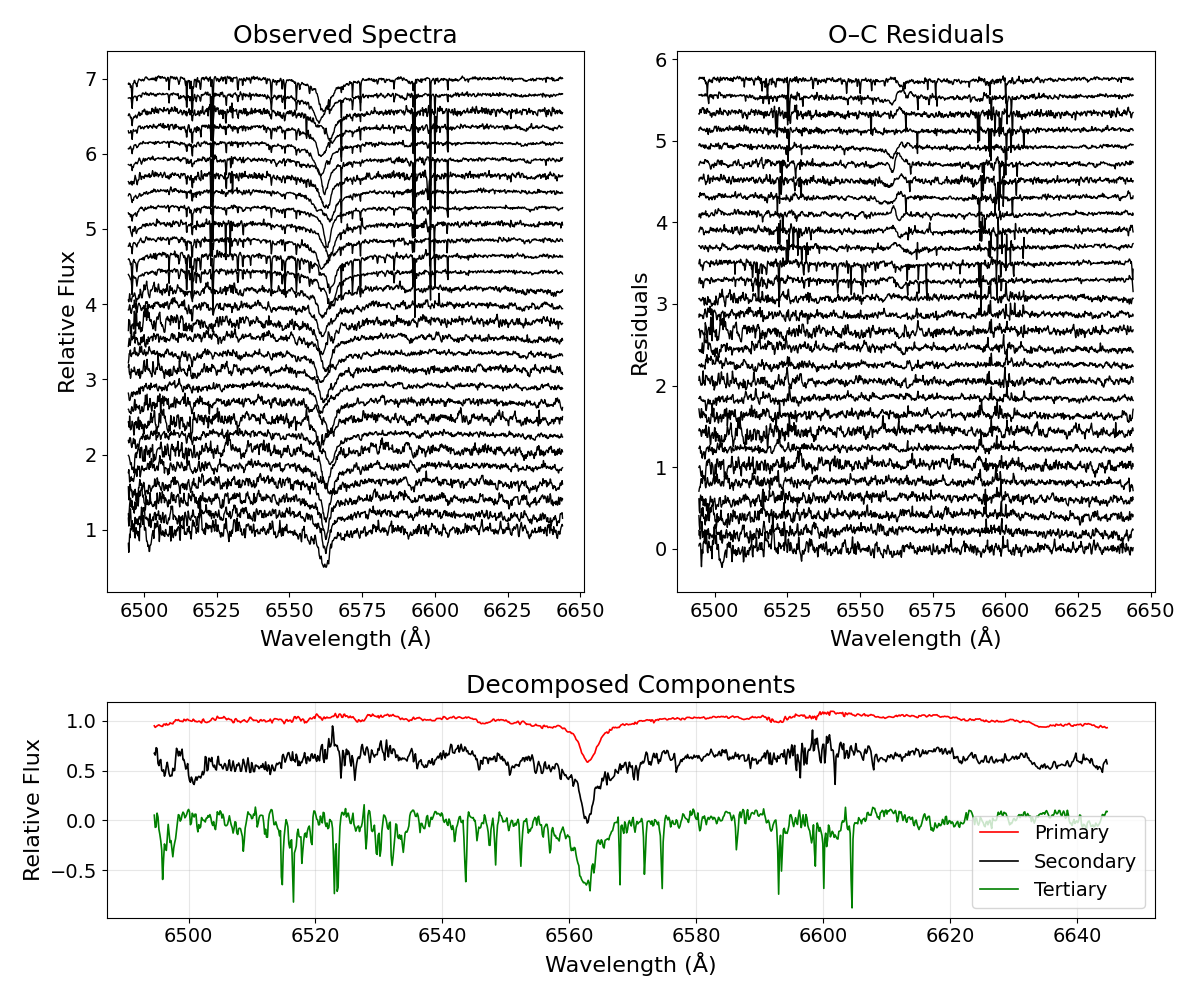}\\[-1ex]
    \includegraphics[width=0.7\linewidth]{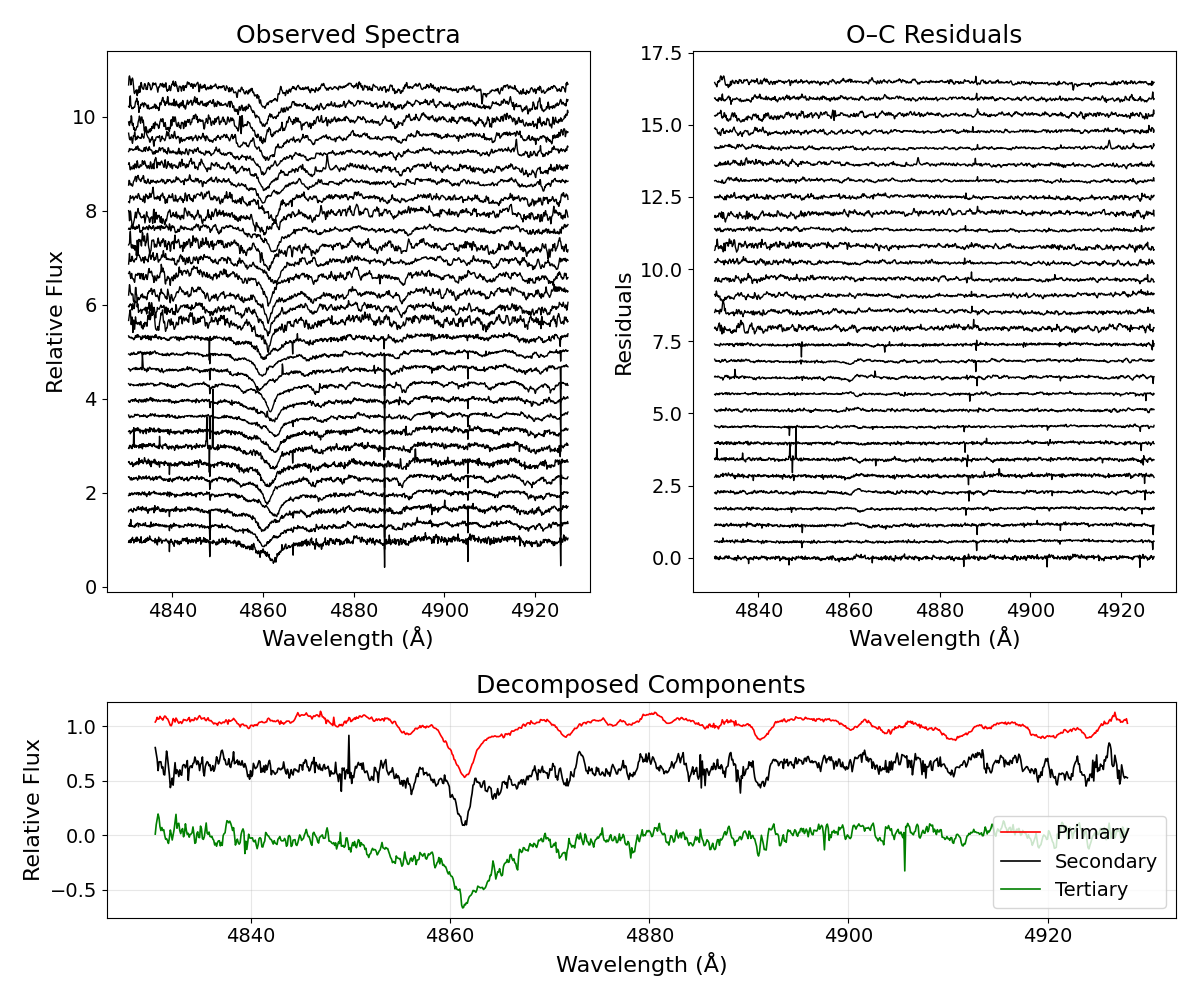}
    \caption{Results of the spectral disentangling analysis of EM\,Boo. The upper panels show the observed composite spectra in the H$_\alpha$ region, the residuals (O–C) of the \texttt{KOREL} fits, and the individual decomposed spectra of the primary, secondary, and tertiary components. The corresponding results for the H$_\beta$ region are presented in the lower panels.}
    \label{fig:tayfbolge}
\end{figure*}

For the EM\,Boo system, the disentangling procedure was performed independently in two spectral regions centred on the H$_\alpha$ and H$_\beta$ lines. The resulting disentangled spectra were subsequently used to determine the atmospheric parameters of the stellar components and to construct synthetic spectra. Figure~\ref{fig:tayfbolge} illustrates, for the H$_\alpha$ region (upper panels), the observed composite spectra, the residuals (O–C) from the $\texttt{KOREL}$ fits, and the decomposed spectra of the three components. The corresponding panels for the H$_\beta$ region are shown in the lower part of the figure. Owing to the significantly larger light contribution of the primary component compared to the secondary and tertiary stars, the disentangled spectrum of the primary is of noticeably higher quality.

Since the effective temperature of the secondary component inferred from the photometric solution is below 10000~K, the model atmosphere code \texttt{ATLAS9}—assuming local thermodynamic equilibrium (LTE)—was adopted for the spectral analysis. Synthetic spectra were computed using the \texttt{SYNTHE} code \citep{Castelli1988, Kurucz1970, Castelli2003}.

The atmospheric parameters of the primary component were determined by comparing its disentangled spectrum with a grid of synthetic models covering effective temperatures in the range $T_{1} = 5000$–$7000$~K (in steps of 100~K) and surface gravities of $\log g_{1} = 3.2$–$3.7$ (in steps of 0.1 dex). The best agreement was obtained for $T_{1} = 6000$~K and $\log g_{1} = 3.5$. For the secondary component, synthetic spectra were generated over the ranges $T_{2} = 5500$–$7000$~K and $\log g_{2} = 3.7$–$4.5$, yielding best-fitting values of $T_{2} = 6350$~K and $\log g_{2} = 4.2$. The projected rotational velocities ($v \sin i$) of primary and secondary components were determined to be $70~\mathrm{km~s^{-1}}$ and $30~\mathrm{km~s^{-1}}$, respectively.
The tertiary component was also successfully disentangled in both spectral regions. However, its atmospheric parameters could not be constrained a priori. Assuming that this component is cooler than 15,000~K, LTE model atmospheres were again adopted, and a wide grid of synthetic spectra was explored, covering effective temperatures of $T_{3} = 5000$–$10,000$~K and surface gravities of $\log g_{3} = 3.5$–$4.5$. The best-fitting model corresponds to $T_{3} = 7000$~K and $\log g_{3} = 4.2$. In addition to atmospheric parameters, the disentangling analysis produces a radial velocity of approximately $30~\mathrm{km~s^{-1}}$ for the tertiary component, further supporting its spectroscopic detection and confirming its contribution to composite spectra. The atmospheric analysis indicates a metallicity of $\mathrm{[M/H]} = -0.1$ dex for the components of the eclipsing binary based on the best match between the profiles of the observed and synthetic metal lines in both spectral regions. For the third component, whose physical association with the binary system remains uncertain, the atmospheric models were computed assuming solar metallicity.

\begin{figure}[!t]
    \centering\includegraphics[width=\linewidth]{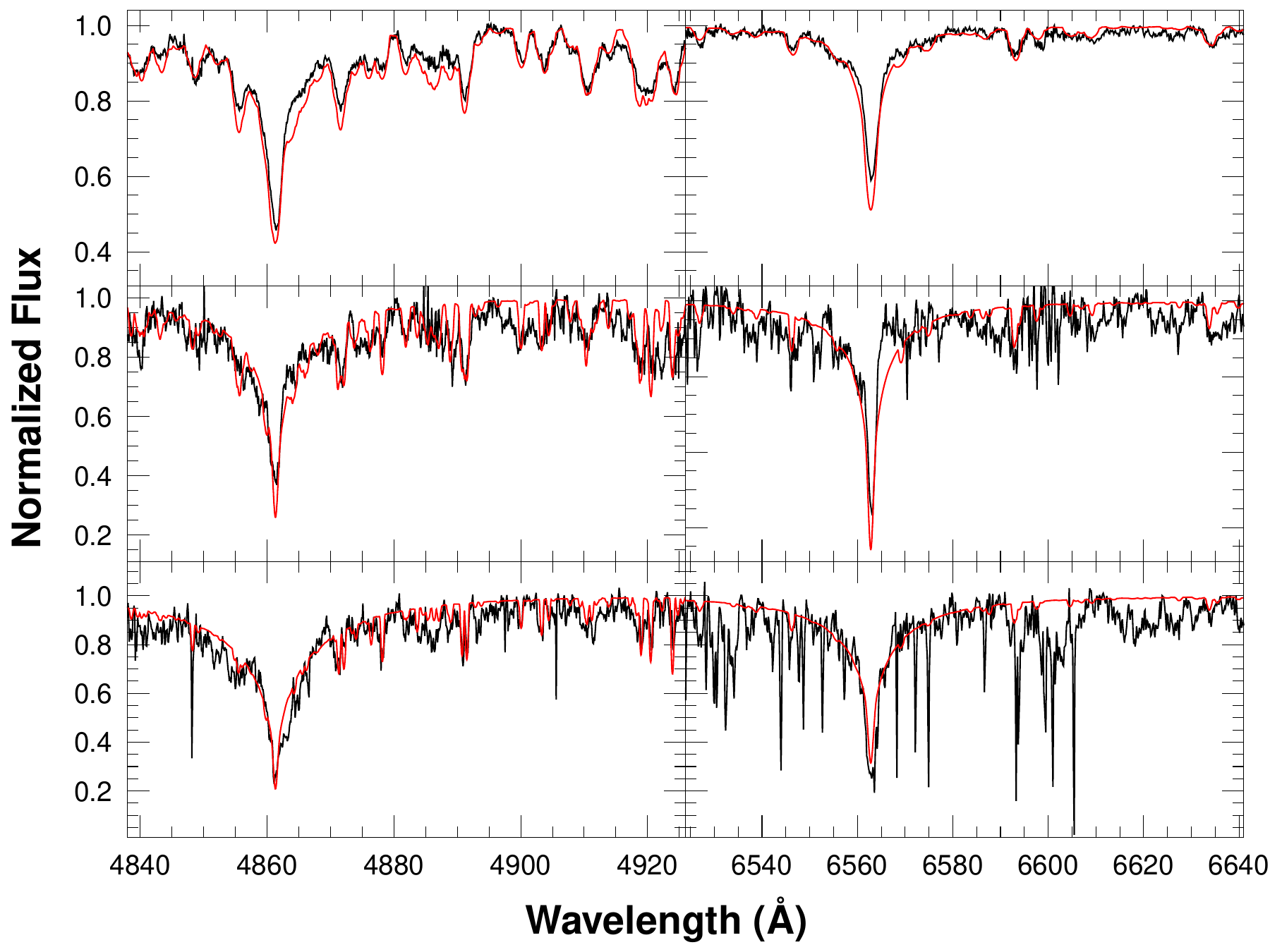}    
    \caption{Disentangled spectra and best-fitting synthetic models of the primary, secondary, and tertiary components of EM\,Boo in the H$_\beta$ (left) and H$_\alpha$ (right) regions. From top to bottom, the panels correspond to the primary, secondary, and tertiary components.}
    \label{fig:tayf_yeni_yeni.pdf}
\end{figure}

Figure~\ref{fig:tayf_yeni_yeni.pdf} presents the disentangled spectra of the primary, secondary, and tertiary components together with their best-fitting synthetic spectra. The left panels show the H$_\beta$ region, while the right panels display the H$_\alpha$ region. In each panel, the spectra and models are shown from top to bottom for the primary, secondary, and tertiary components, respectively. 

The third component detected in both the photometric and spectroscopic data contributes significantly 
to the observed flux of the system. While its spectral features and radial velocity ($\sim$30 km s$^{-1}$) confirm its presence, the available observations are insufficient to determine whether it is physically bound to the eclipsing pair. Notably, the astrometric parameters reported in WDS show a gradual change in separation and position angle between 1968 and 2014, with the \textit{HIPPARCOS} epoch lying midway between them. This trend may hint at a long-term orbital motion involving the tertiary component, but further long-term spectroscopic or 
astrometric observations are required to resolve this issue.

\section{Astrophysical parameters and the distance of the system}\label{sec:section6}
\subsection{Fundamental stellar parameters}

The fundamental astrophysical parameters of EM\,Boo derived from the combined photometric and spectroscopic analysis are listed in Table~\ref{tab:tess_params}. The precisely determined masses and radii provide strong constraints on the physical properties and evolutionary status of both stellar components.

The effective temperatures of the primary and secondary stars were obtained from atmospheric modelling of the disentangled spectra and are found to be $T_{\mathrm{eff}} = 6000 \pm 100$~K and $6350 \pm 100$~K, respectively. Based on these emperatures, together with the derived surface gravities, the spectral types were assigned as F9–G0\,III/IV for the primary component and F6\,V for the secondary. These classifications are consistent with the inferred luminosities and the positions of the stars in the mass--radius plane.

The primary component has a mass of $1.48\,M_{\odot}$ and a radius of $3.21\,R_{\odot}$, indicating that it has evolved off the main sequence and is currently in a subgiant or early giant evolutionary phase. This interpretation is supported by its relatively low surface gravity ($\log g = 3.59$) and high luminosity ($\log (L/L_{\odot}) = 1.08$). In contrast, the secondary star, with a mass of $1.20\,M_{\odot}$, a radius of $1.41\,R_{\odot}$, and a surface gravity of 
$\log g = 4.22$, is consistent with a main-sequence evolutionary status.

The projected rotational velocities derived from the atmospheric analysis are consistent with the computed synchronous rotational velocities within the uncertainties, suggesting that both components are synchronised with the orbital motion. This behaviour is expected given the relatively small orbital separation of the system.
Overall, the derived fundamental parameters present a coherent picture of an evolved primary star accompanied by a less evolved main-sequence secondary, in agreement with predictions from binary stellar evolution theory.
\begin{table*}[!t]
\centering
\caption{Astrophysical parameters of EM\,Boo.}
\label{tab:tess_params}
\begin{tabular}{lccc}
\hline
Parameter & Symbol & Primary & Secondary \\
\hline
Spectral type & Sp & F9.5III/IV  & F6 V \\
Mass ($M_{\odot}$) & $M$ & $1.476 \pm 0.097$ & $1.195 \pm 0.082$ \\
Radius ($R_{\odot}$) & $R$ & $3.213 \pm 0.063$ & $1.407 \pm 0.030$ \\
Separation ($R_{\odot}$) & $a$ & \multicolumn{2}{c}{$10.6 \pm 0.2$} \\
Surface gravity (cgs) & $\log{g}$ & $3.593 \pm 0.047$ & $4.219 \pm 0.050$ \\
Integrated visual magnitude (mag) & $V$ & \multicolumn{2}{c}{9.06}\\
Individual visual magnitudes (mag) & $V$ & $9.61 \pm 0.04$ & $11.14 \pm 0.09$ \\
Effective temperature (K) & $T_{\mathrm{eff}}$ & $6000 \pm 100$ & $6350 \pm 100$ \\
Luminosity ($L_{\odot}$) & $\log L$ & $1.08 \pm 0.04 $ & $0.47 \pm 0.04$ \\
Bolometric magnitude (mag) & $M_{\mathrm{Bol}}$ & $2.03 \pm 0.10$ & $3.58 \pm 0.11$ \\
Absolute visual magnitude (mag) & $M_{\rm V}$ & $2.05 \pm 0.08$ & $3.63 \pm 0.10$ \\
Bolometric correction (mag) & BC & $-0.05 \pm 0.01$ & $-0.01 \pm 0.01$ \\
Computed synchronization velocity (km\,s$^{-1}$) & $v_{\mathrm{synch}}$ & $ 67\pm1 $ & $ 29\pm1 $ \\
Projected rotational velocity (km\,s$^{-1}$) & $v \sin i$ & $70 \pm 10 $ & $30 \pm 10$ \\
\textit{Gaia} distance (pc) & $d_{\rm Gaia}$ & \multicolumn{2}{c}{$252\pm 11$}\\
Photometric distance (pc) & $d_{\rm ph}$ & \multicolumn{2}{c}{$300\pm 10$}\\
SED distance (pc) & $d_{\rm SED}$ & \multicolumn{2}{c}{$313\pm 25$}\\
\hline
\end{tabular}
\end{table*}

\subsection{Astrometric and SED-based Distance Estimates}

The \textit{HIPPARCOS} measured the trigonometric parallax of EM\,Boo as $\varpi_{\rm Hip}=1.33\pm1.45$ mas \citep{vanLeuwen2007} while the Global Astrometric Interferometer for Astrophysics satellite \citep[\textit{Gaia};][]{Gaia2016} gives the parallax as $\varpi_{\rm Gaia}=3.9699\pm0.1812$ mas in its third data release. There is a clear discrepancy between these two measurements. Such discrepancies between \textit{HIPPARCOS} and \textit{Gaia} trigonometric parallaxes are not uncommon in multiple or interacting stellar systems. In these cases, the orbital motion of the components can shift the photocenter of the system, leading to biased astrometric solutions when a single-star model is assumed. \textit{HIPPARCOS}, with its more limited angular resolution, was particularly prone to such effects in close binaries \citep{vanLeuwen2007}. \textit{Gaia} DR3 has significantly improved astrometric precision, but systematic errors may still arise in non-single stars due to unresolved orbital motion or variability \citep{Lindegren2021, GaiaCollaboration2023}. The large difference between the \textit{HIPPARCOS} trigonometric parallax ($1.33\pm1.45$ mas) and the \textit{Gaia} DR3 value ($3.9699\pm0.1812$ mas) for EM\,Boo is therefore most likely related to the system’s multiplicity and the limitations of single-star astrometric solutions. This highlights the importance of combining trigonometric parallaxes with independent distance estimates from photometric and spectroscopic analyses. 

One such approach is the photometric distance derived from using the component's absolute magnitudes and visual brightness in the distance modulus. In this method, the interstellar extinction plays a key role for estimating a reliable distance. Interstellar extinction can be estimated from the modelling of multi-wavelength flux data (SED) of the system, as described by \citet{Bakis2022} and \citet{Eker2023}, who successfully applied this approach to double-lined eclipsing binary stars. 

\begin{figure*}[!t]
    \centering
    \includegraphics[width=0.99\linewidth]{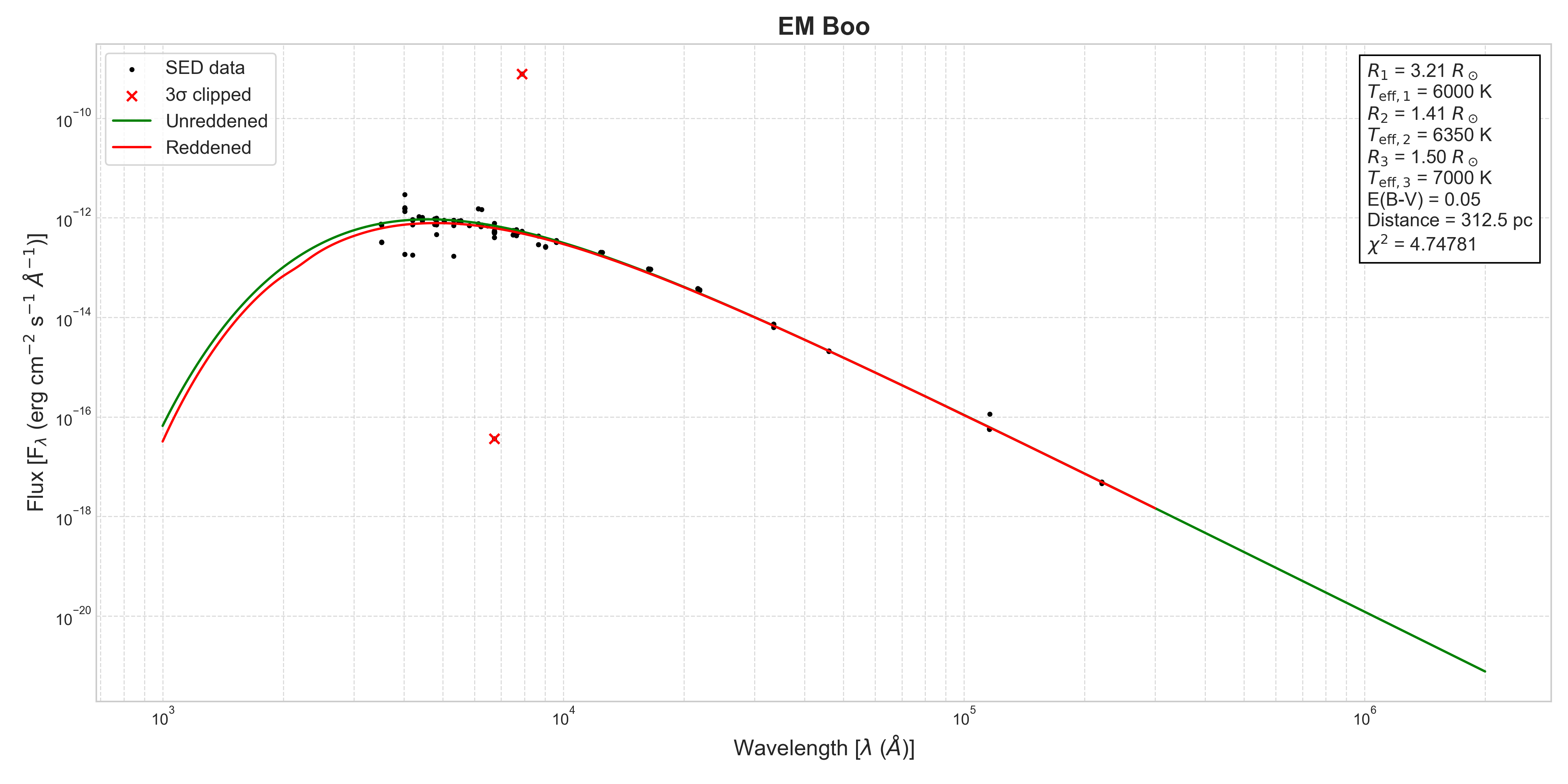}\vspace*{-10pt}
    \caption{Best-fitting SED model of EM\,Boo, showing the contributions of the primary, secondary, and tertiary components.}
    \label{fig:sed}
\end{figure*}

Here, we collected the SED data of EM\,Boo from the SIMBAD \citep{Wenger2000} database and fitted three components in the system using the radius and temperature information given in Table~\ref{tab:tess_params}. For the tertiary component, the radius was adopted as a representative value ($1.5\,R_\odot$) for a 7000 K main-sequence star, consistent with standard mass–radius relations given in \citet{Eker2018}. The colour excess $E(B-V)$ and distance $d_{\rm SED}$ were iteratively adjusted to obtain the best-fitting SED model. The resulting fit, shown in Figure~\ref{fig:sed}, yields $E(B-V)=0.05$ mag and a parallax $\varpi_{\rm SED}=3.2\pm0.23$ mas, corresponding to a distance of $d_{\rm SED}=313\pm25$ pc.

The SED-based distance of 313 pc lies between the discrepant \textit{HIPPARCOS} and \textit{Gaia} values, supporting the \textit{Gaia} DR3 parallax as the more reliable estimate for EM\,Boo.

Finally, the bolometric correction values derived from the components' temperatures were adopted as defined in the study by \cite{Yucel2026AJ}. Accordingly, the corrections are \(BC_{1} = -0.05 \, \mathrm{mag}\) and \(BC_{2} = -0.01 \, \mathrm{mag}\). From the bolometric absolute magnitudes \(M_{\mathrm{bol},1} = 2.03 \, \mathrm{mag}\) and \(M_{\mathrm{bol},2} = 3.58 \, \mathrm{mag}\), the visual absolute magnitudes of each component were determined as \(M_{\rm V,1} = 2.08 \, \mathrm{mag}\) and \(M_{\rm V,2} = 3.59 \, \mathrm{mag}\). 

In the distance modulus, if we adopt the reddening value from the SED analysis \(E(B-V) = 0.05 \, \mathrm{mag}\), which corresponds to \(A_{\rm V} = 0.155 \, \mathrm{mag}\), and calculate the apparent visual magnitudes from the light contributions of the components, the distances are found to be \(d_{1} = 299 \, \mathrm{pc}\) for the first component and \(d_{2} = 301 \, \mathrm{pc}\) for the second component.  
We note that the radius of the tertiary component cannot be constrained directly from the light-curve and radial-velocity analysis, and is therefore adopted as $R_3 = 1.5\,R_{\odot}$ based on its effective temperature. This assumption affects the absolute flux scaling of the SED model. For a fixed effective temperature, the flux contribution scales approximately as $F_{\lambda} \propto R^2/d^2$, implying that the distance associated with the tertiary component scales nearly linearly with the adopted radius. As a result, a variation of $R_3$ by $\pm 20\%$ would lead to a comparable change in the inferred distance from the tertiary contribution, although the overall effect is mitigated by the dominant flux contribution of the eclipsing binary components.

Considering the distance derived from the SED analysis \((d_{\mathrm{SED}} = 313 \, \mathrm{pc})\) and the photometric distance obtained from the absolute magnitudes of the components \((d_{\mathrm{ph}} = 300 \, \mathrm{pc})\), even within this level of uncertainty, the comparison suggests that the distance determined by the \textit{Gaia} satellite may be underestimated.

\section{Evolutionary Status}\label{sec:evolutionary_status}

The evolutionary calculations were performed using \texttt{MESA} version 23.05.1 \citep{Paxton2011, Paxton2013, Paxton2015, Paxton2018, Paxton2019, Jermyn2023}. Using its \texttt{binary} module, \texttt{MESA} enables the modeling of the simultaneous evolution of both stars in a binary system, including the calculation of orbital parameters that play a critical role in determining the time and condition of mass transfer.

Our initial analysis focused on locating the components of EM\,Boo on the HR diagram in order to test the consistency of our results with \texttt{MESA} evolutionary models. The metallicity analysis indicates that both components metallicities are $-0.1$ dex in terms of Solar. We have adopted the solar composition of \citet{Grevesse1998}, with $Z_{\odot}=0.0169$, which corresponds to $0.0134$ for both components in our evolutionary calculations. In \texttt{MESA}, the helium abundance is given by the relation $(Y=0.24+2\times Z)$. Therefore, adopted values for $Y$ are 0.267 and 0.267 for both components by default on \texttt{MESA}. Another key parameter in stellar modeling is the mixing length parameter ($\alpha_{\rm MLT}$), which describes convective energy transport, which was represented the characteristic distance, expressed in units of pressure scale height, that a convective element travels before dissolving \citep[see][]{Joyce2023}. \texttt{MESA} adopts mixing length parameter as 2.0 by default. 

\begin{figure}[!t]
    \centering
    \includegraphics[width=0.99\linewidth]{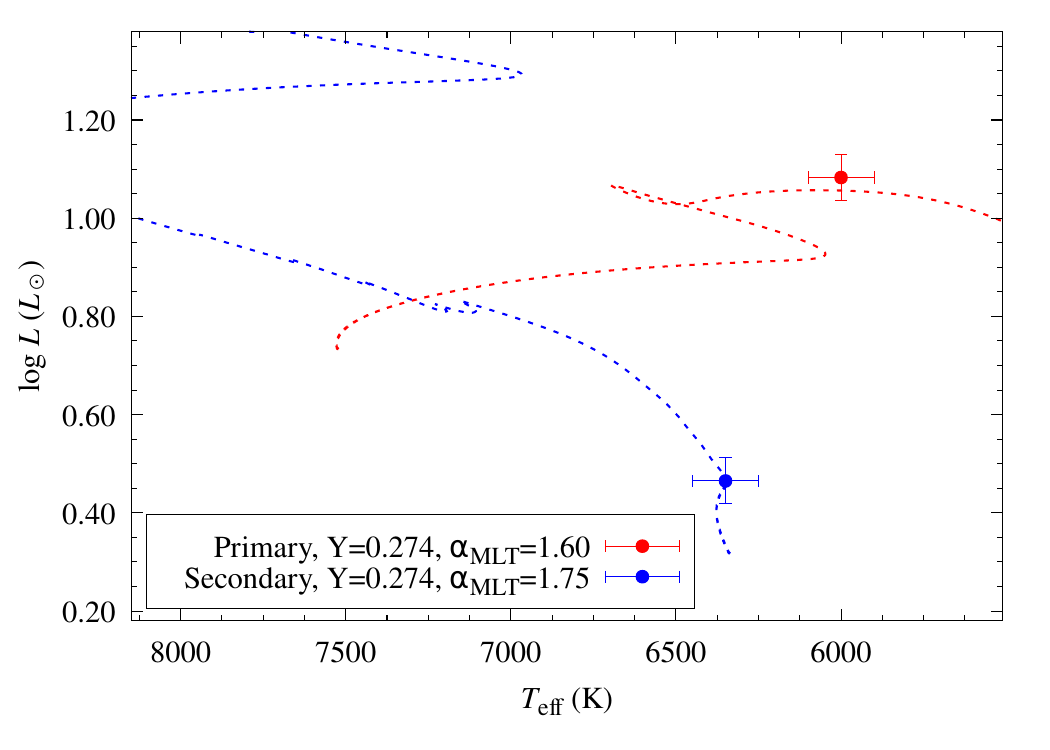}\\[-2ex]
    \includegraphics[width=0.99\linewidth]{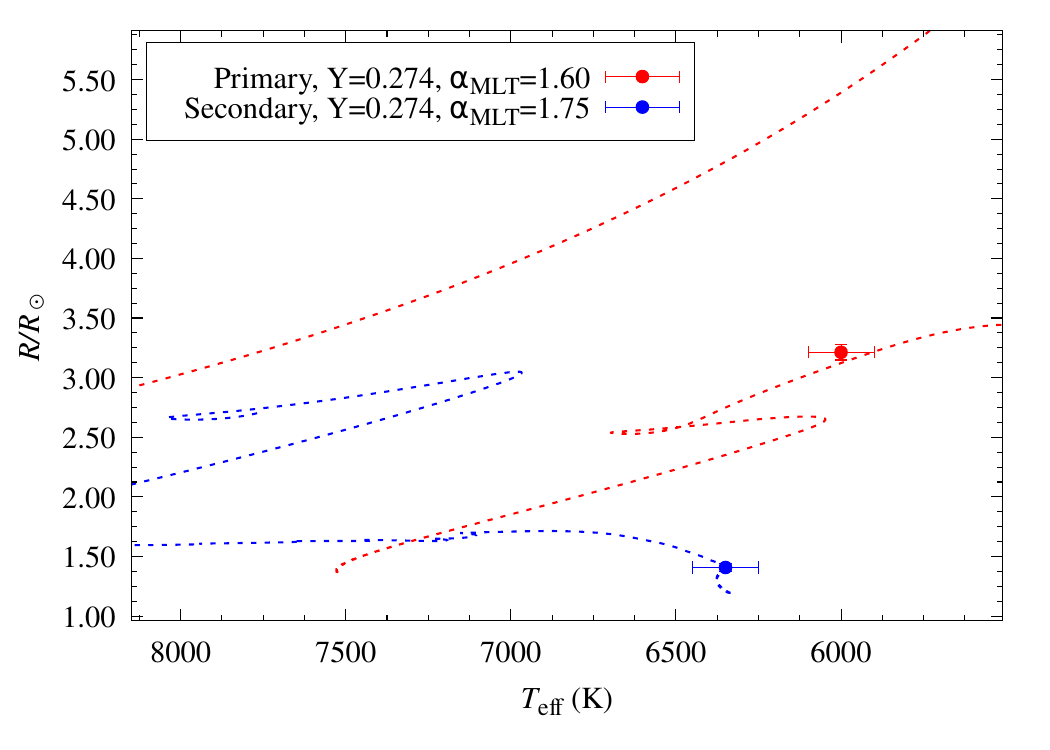}\\[-2ex]
    \includegraphics[width=0.99\linewidth]{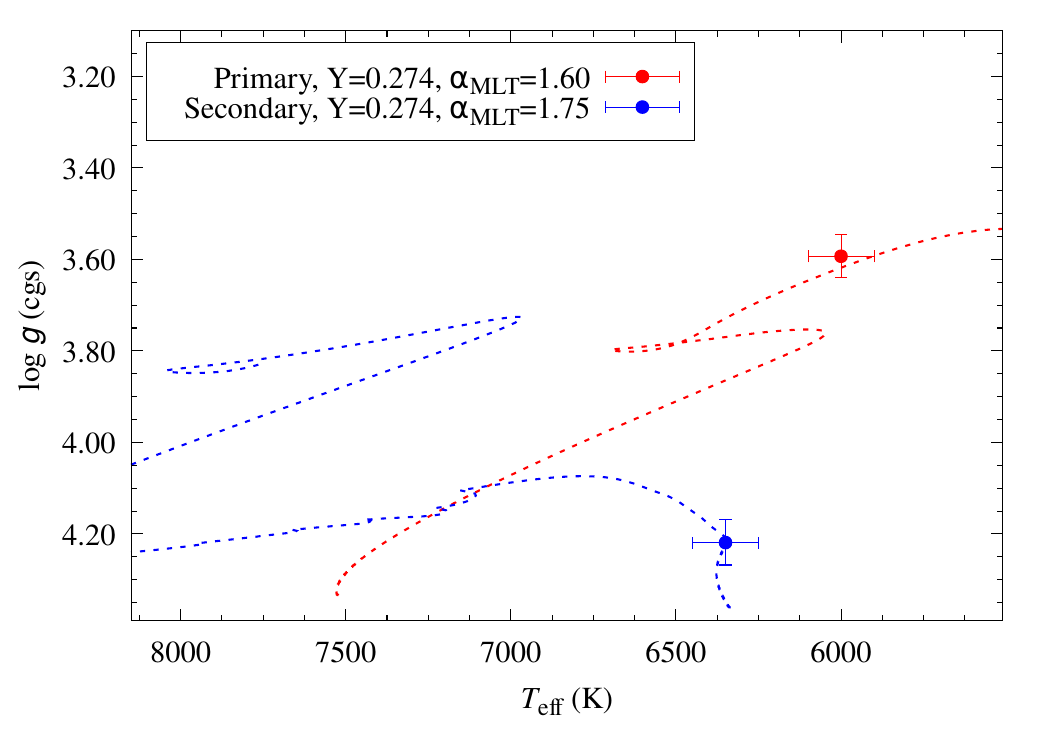}\\[-2ex]
    \caption{The evolutionary tracks of the primary and secondary components of EM\,Boo are shown in the $\log (L/L_{\odot})$–$T_{\rm eff}$ (left), $R/R_{\odot}$–$T_{\rm eff}$ (middle), and $\log g$–$T_{\rm eff}$ (right) diagrams, respectively.}
    \label{fig:hr}
\end{figure}

The important part in stellar evolution calculations is that every star is unique. Thus, the general models cannot be adopted to every star, especially helium abundances and $\alpha_{\rm MLT}$, which drastically change stellar interiors. In this context, we treated $Y$ and $\alpha_{\rm MLT}$ as free parameters in our calculations \citep[e.g.][]{Fernandes2012, Valle2014, Verma2022, Yucel2026}. We have investigated a range starting just below the primordial helium abundance $Y=0.245$ \citep{Peimbert2007} and extending up to $Y=0.291$, with a step size of 0.002 \citep{Lebreton2014}. $\alpha_{\rm MLT}$ was also evaluated from 1.2 to 2.0 with 0.05 steps. The final adopted values of $Y$ and $\alpha_{\rm MLT}$ were chosen to reproduce the observed positions of the stars in the HR diagrams (see Figure~\ref{fig:hr}). The analysis yielded an age of $2.38_{-0.02}^{+0.02}$ Gyr for the primary star ($Y = 0.274$, $\alpha_{\rm MLT}= 1.60$) and $2.38_{-0.1}^{+0.1}$ Gyr for the secondary star ($Y = 0.274$, $\alpha_{\rm MLT}=1.75$). Taking the weighted mean of the two components, the system age was determined as $t = 2.38\pm 0.05$ Gyr.

The analysis shows that EM\,Boo is a detached binary system, with the primary component is at the shell-H burning phase while the secondary component is still in the main-sequence (MS) phase. This implies that no mass transfer has occurred since the components entered the MS. Consequently, the present-day masses of the stars are essentially the same as their initial values at formation. The only parameters that have evolved since formation are the orbital elements, period and eccentricity, besides stars themselves. Given that our RV-LC analysis yielded zero eccentricity, it is not possible to determine when the orbit of EM\,Boo was circularized. Therefore, in our evolutionary calculations, we assumed that the system was formed in a circular orbit.

\begin{figure}[!t]
    \centering
    \includegraphics[width=0.99\linewidth]{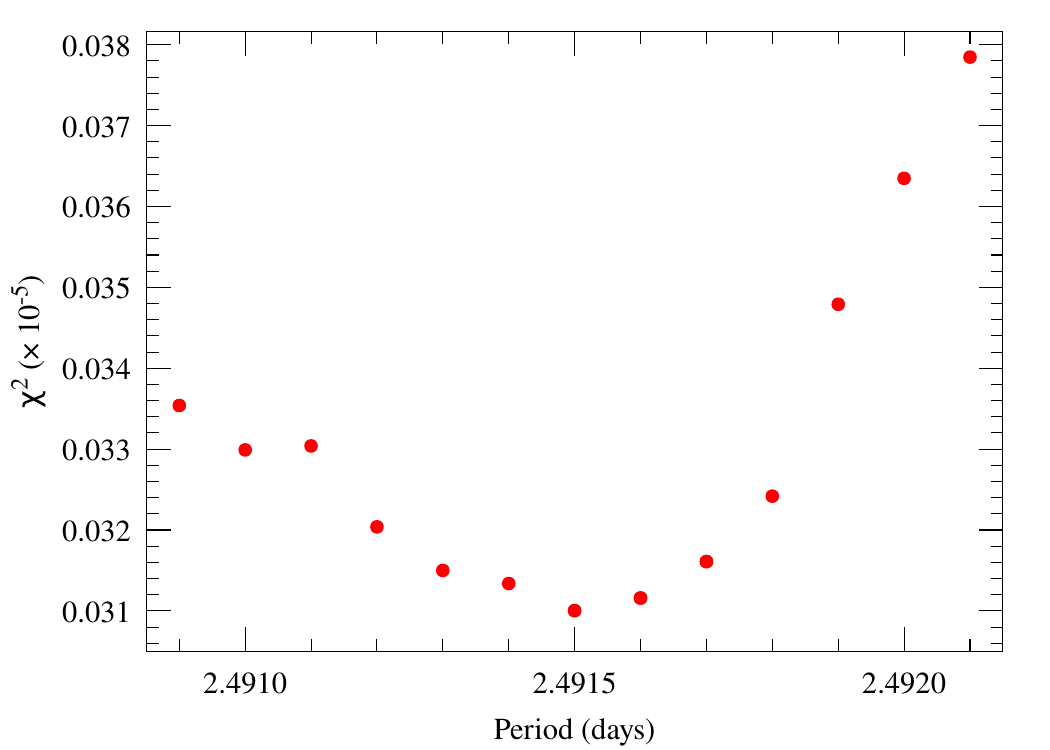}
    \caption{The result of the grid search yielded the best initial orbital condition, with an orbital period of $P=2.4915$ days. Goodness of the fit ($\chi^2$) is plotted against the tested period values, with the minimum $\chi^2$ indicating the best fit period.}
    \label{fig:period}
\end{figure}

\begin{figure}[!t]
    \centering
    \includegraphics[width=0.99\linewidth]{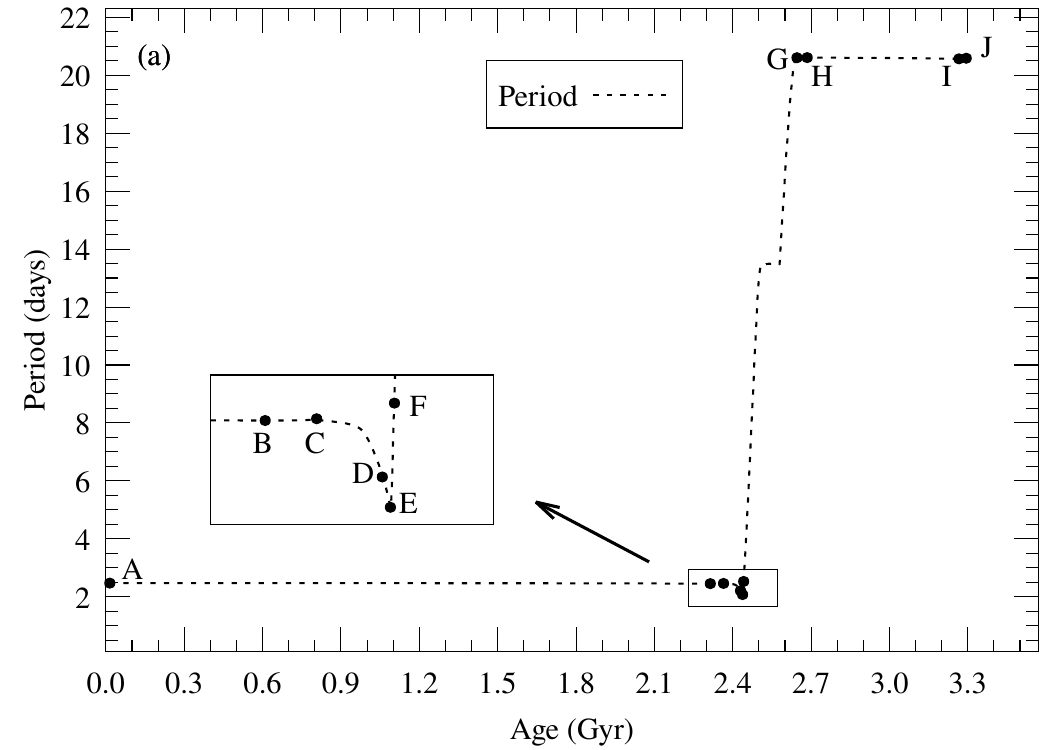}\\[5pt]
    \includegraphics[width=0.99\linewidth]{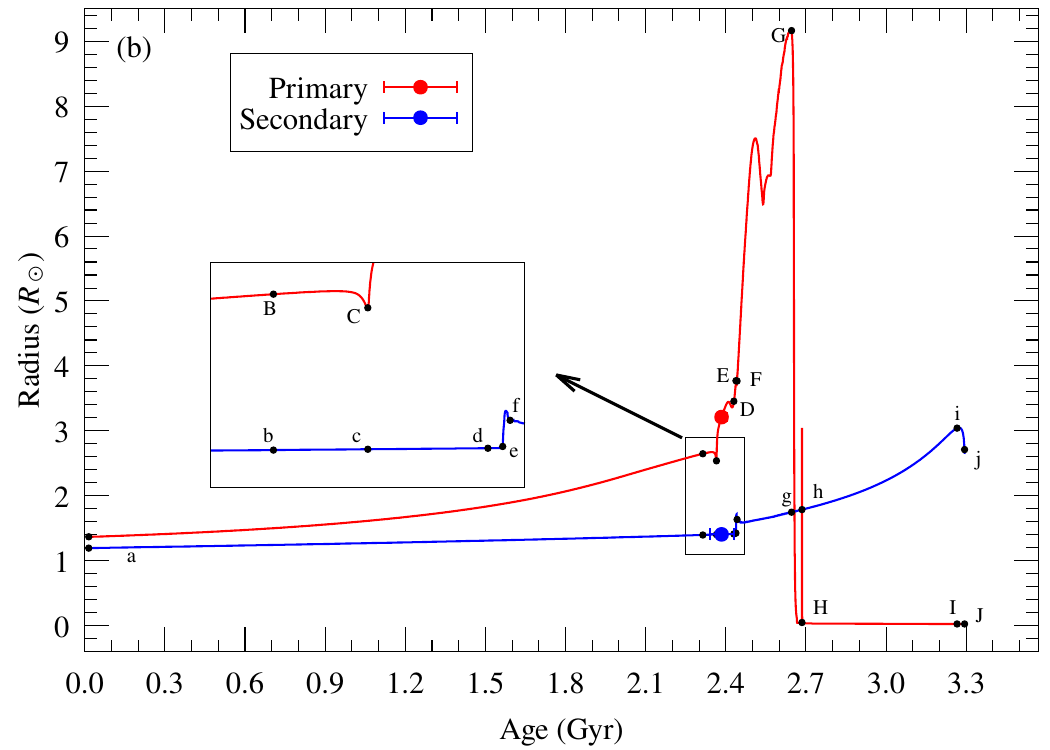}
    \caption{Change of orbital period (a), and radius (b) of the components of EM\,Boo with time.}
    \label{fig:radius}
\end{figure}

\begin{figure}[!t]
    \centering
    \includegraphics[width=0.99\linewidth]{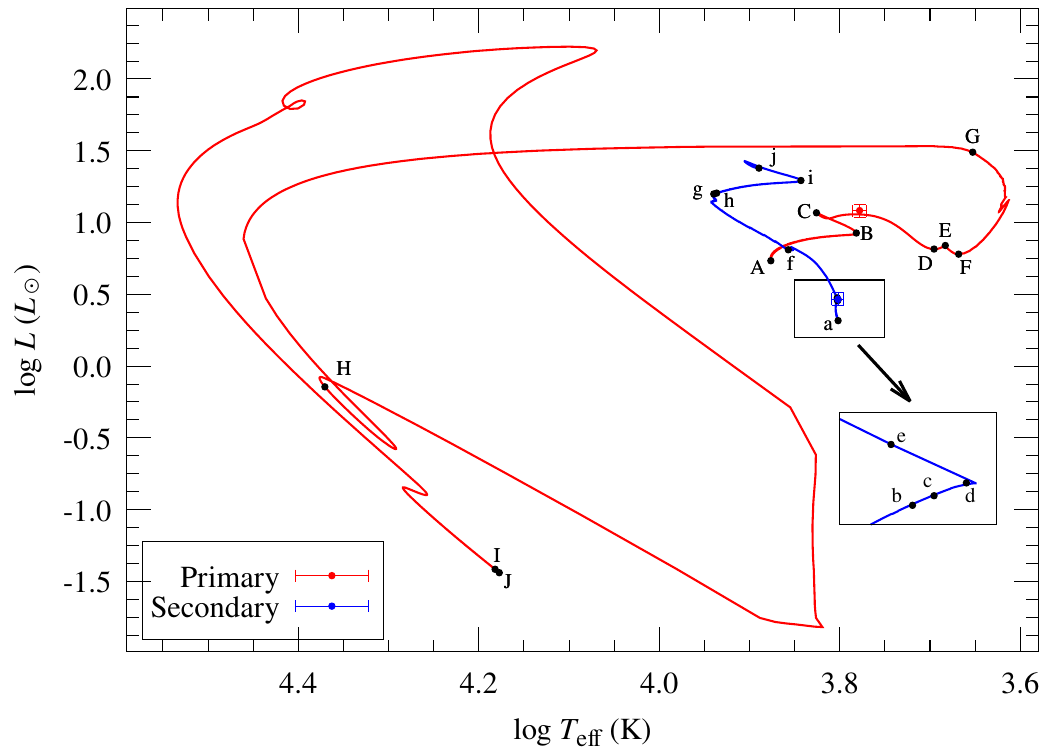}
    \caption{\texttt{MESA} evolutionary tracks of the primary and secondary components of EM\,Boo displayed in the $\log (L/L_{\odot})\times T_{\rm eff}$ diagram, showing their positions and evolutionary stages.}
    \label{fig:HR}
\end{figure}

After locating the components of EM\,Boo on the Hertzsprung-Russell (HR) diagram, the next step was to determine the initial orbital conditions. To this end, we performed a grid search, a method widely used in the literature \citep[e.g.,][]{Rosales, Soydugan2020, Yucel2022, Yucel2024, Yucel2025}. In this approach, evolutionary calculations were carried out starting from a randomly selected initial orbital period and continued until the present-day period, $2.44623996$ was reached. A $\chi^2$ analysis was then performed using the current radii and temperatures of the components. The best-fit model corresponded to an initial orbital period of $2.4915$ days (Figure~\ref{fig:period}). Since the components of EM\,Boo are low-mass components, angular momentum loss via magnetic braking can drive orbital decay and synchronization on timescales that depend sensitively on stellar magnetic field strength and wind properties \citep{Skumanich1972, Verbunt1981, Rappaport1983}. In our calculations, we accounted for magnetic braking, given that both components possess convective envelopes \citep{Rappaport1983}. Tidal dissipation acts to synchronize rotation and circularize the orbit on characteristic timescales depending on the stellar structure and separation \citep{Zahn1977}. For tidal synchronization, we adopted the “Orb period” option, which enforces synchronization on the timescale of the orbital period. Roche lobe radii were computed following the prescription of \citet{Eggleton1983}, while mass transfer rates in Roche lobe–overflowing binaries were calculated according to \texttt{roche\_lobe} setting of \texttt{MESA}, which assumes components are staying in their Roche lobe during the mass transfer. 
After determining the initial orbital period, we carried out new evolutionary calculations starting from the derived value of 2.4915 days and continued them until the secondary component reached the Terminal-Age Main Sequence (TAMS) phase. The mean mass-transfer efficiency coefficients were adopted following \citet{Paxton2015}, \citet{Rosales}, and \citet{Soydugan2020}, with values of 0.4, 0.1, and 0.1 for $\alpha$, $\beta$, and $\gamma$, respectively. Finally, our calculations indicate that the system is at the edge of mass transfer, which will probably start in about 60 Myr. The variations in orbital period and component radii throughout the evolution are shown in Figure~\ref{fig:radius}. A more detailed evolutionary track of both components, including characteristic timescales, is provided in Table~\ref{tab:timestamp} and shown in Figure~\ref{fig:HR}.

\begin{table*}[ht]
   \centering
\setlength{\tabcolsep}{4.5pt}
\renewcommand{\arraystretch}{1.1}
{\scriptsize
\centering
    \caption{Detailed evolution of EM\,Boo with time-stamps.}
    \label{tab:timestamp}
        \begin{tabular}{llcccccccccccc}
    \toprule\noalign{\vspace*{-2pt}}
        Mark & \multirow{2}*{Evolutionary Status} & Age & $P$  & \multicolumn{5}{c}{Primary} & \multicolumn{5}{c}{Secondary} \\
       \cline{5-10} \cline{11-14}\noalign{\vspace*{2pt}}
         (Pri/Sec) & & (Gyr) & (day) & $T_\mathrm{eff}$ (K) & $\log L$ ($L_\odot$)  & $M$ ($M_\odot$) & $R$ ($R_\odot$) & $\log{g}$ (cgs)& $T_\mathrm{eff}$ (K) & $\log L$ ($L_\odot$) & $M$ ($M_\odot$) & $R$ ($R_\odot$) & $\log{g}$ (cgs)\\[-1ex]
        \hline
        A/a & ZAMS & 0.015 & 2.471 & 7523 & 0.734 & 1.476 & 1.370 & 4.334 & 6336 & 0.317 & 1.195 & 1.196 & 4.360 \\ 	
        B/b & Core contraction & 2.314 & 2.452 & 6048 & 0.927 & 1.476 & 2.649 & 3.761 & 6347 & 0.456 & 1.195 & 1.398 & 4.224 \\
        C/c & TAMS & 2.365 & 2.460 & 6696 & 1.068 & 1.476 & 2.540 & 3.798 & 6343 & 0.459 & 1.195 & 1.404 & 4.220 \\
        D/d & RGB & 2.430 & 2.207 & 4961 & 0.815 & 1.476 & 3.457 &3.530 & 6337 & 0.463 & 1.195 & 1.413 & 4.215\\
        E/e & Mass Transfer & 2.438 & 2.075 & 4820 & 0.840 & 1.475 & 3.773 & 3.454 & 6351 & 0.475 & 1.195 & 1.427 & 4.207 \\
        F/f & Re-RGB & 2.442 & 2.528 & 4659 & 0.780 & 1.100 & 3.767 & 3.327 & 7194 & 0.810 & 1.383 & 1.637 & 4.151 \\
        G/g & MT Finish & 2.646 & 20.605 & 4495 & 1.490 & 0.260 & 9.161 & 1.930 & 8702 & 1.199 & 1.803 & 1.749 & 4.208 \\
        H/h & H Shell Flash & 2.685 & 20.609 & 23466 & -0.145 & 0.260 & 0.051 & 6.435 & 8638 & 1.205 & 1.803 & 1.788 & 4.189 \\
        I/i & Sec. Core Contr. & 3.266 & 20.567 & 15202 & -0.142 & 0.260 & 0.028 & 6.951 & 6967 & 1.293 & 1.803 & 3.042 & 3.728 \\
        J/j & Sec. TAMS & 3.294 & 20.587 & 15039 & -0.144 & 0.260 & 0.028 & 6.956 & 7754 & 1.379 & 1.803 & 2.712 & 3.828 \\
        \bottomrule
    \end{tabular} 
    }
    \end{table*}


\section{Concluding Remarks}\label{sec:conclusions}
In this study, we carried out a comprehensive photometric, spectroscopic, and evolutionary investigation of the triple star system EM\,Boo by combining AUT25 and high-precision \textit{TESS} photometry with spectroscopic data obtained from the UBT60 telescope and the ELODIE archive. Through a stepwise analysis strategy, we disentangled the complex contributions of the individual components and derived a self-consistent set of stellar and orbital parameters. The analysis of the light curves enabled the determination of phase-dependent light contributions, which were subsequently incorporated into the spectroscopic analysis to reliably extract the radial velocity curves and obtain disentangled spectra of all three components. Notably, this study represents the first successful disentangling of the tertiary spectrum in EM\,Boo, establishing a benchmark for future analyses of hierarchical triple systems.

The atmospheric analysis of the disentangled spectra reveals that the components of EM\,Boo occupy clearly distinct evolutionary stages. The primary star has evolved off the main sequence and is currently in a subgiant or early giant phase, whereas the secondary component remains on the main sequence. For the tertiary component, the atmospheric modeling of its disentangled spectrum yields an effective temperature of $T_{\mathrm{eff}}=7000$~K and a surface gravity of $\log g = 4.1$, consistent with a main-sequence star of intermediate spectral type. This interpretation is supported by the locations of all three stars in the mass--radius plane, their inferred luminosities, and the derived surface gravities, which are in agreement with the assigned spectral classifications.

Stellar evolution models computed with the MESA code further confirm that EM\,Boo is a detached system in which the observed differences between the components arise from their intrinsic stellar evolution rather than from binary interaction effects. The absence of mass transfer signatures indicates that the present-day stellar parameters closely reflect the initial conditions at formation, making EM\,Boo an excellent laboratory for testing stellar evolution in multiple systems.

The distance to EM\,Boo was independently estimated using both photometric methods and SED analysis. The two approaches yield mutually consistent results, indicating a robust distance determination. Both methods suggest a slightly larger distance than that inferred from \textit{Gaia} trigonometric parallaxes, pointing to possible limitations in the current astrometric solution for this system.

Overall, this study provides one of the most detailed and self-consistent characterizations of the EM\,Boo system to date. The combination of phase-dependent photometric analysis, spectroscopic disentangling, atmospheric modeling, and evolutionary calculations demonstrates the effectiveness of a multi-step approach for studying complex multiple stellar systems. The present analysis offers a robust picture of the system. Future high signal-to-noise and long-term spectroscopic observations, together with improved astrometric constraints from upcoming \textit{Gaia} data releases, will refine the properties of the tertiary component and better constrain the global architecture of EM\,Boo.

\vspace{5pt}
\begin{description}
  \item[Peer Review:] Externally peer-reviewed.
  \item[Author Contribution:] Conception/Design of study - H.B., Ö.H.Y.; Data Acquisition - Ö.H.Y.; Data Analysis/Interpretation - Ö.H.Y. B.Ö., H.B.; ~Drafting Manuscript - H.B.;~Critical Revision of Manuscript - H.B., Ö.H.Y;~Final Approval and Accountability - H.B.; ~Supervision: H.B.
\item[Conflict of Interest:] Author declared no conflict of interest.
  \item[Financial Disclosure:] ---
\end{description}
\vspace{5pt}

\section*{Acknowledgments}
The authors would like to thank Prof. Volkan Bakış and Dr. Gökhan Yücel for their valuable contributions, insightful comments, and support of this study on EM Boo, as well as the anonymous referees for their constructive comments, which significantly improved the manuscript. The authors also thank the Department of Space Sciences and Technologies at Akdeniz University for granting access to the 60-cm and 25-cm telescopes, which were used to obtain part of the photometric and spectroscopic observations presented in this work. This research uses data obtained from the ELODIE archive at Observatoire de Haute-Provence (OHP). This study also utilized publicly available data from the Transiting Exoplanet Survey Satellite (\textit{TESS}), retrieved from the Mikulski Archive for Space Telescopes (MAST). This work has made use of data from the European Space Agency (ESA) mission {\it Gaia} (\url{https://www.cosmos.esa.int/gaia}), processed by the {\it Gaia} Data Processing and Analysis Consortium (DPAC, \url{https://www.cosmos.esa.int/web/gaia/dpac/consortium}). Funding for the DPAC has been provided by national institutions, in particular the institutions participating in the {\it Gaia} Multilateral Agreement. This research has made use of the SIMBAD database, operated at CDS, Strasbourg, France. This research has made use of the Astrophysics Data System, funded by NASA under Cooperative Agreement 80NSSC21M0056.

\spacebref{5pt}{-25pt}
\bibliographystyle{mnras}
\bibliography{reference}

\bsp	
\label{lastpage}
\end{document}